\documentclass[journal]{IEEEtran}
\usepackage{amsmath,amsfonts,amssymb}
\usepackage{algorithmic}
\usepackage{algorithm}
\usepackage{array}
\usepackage[caption=false,font=normalsize,labelfont=sf,textfont=sf]{subfig}
\usepackage{bm} 
\usepackage{mathtools} 
\usepackage{textcomp}
\usepackage{url}
\usepackage{verbatim}
\usepackage{enumitem}
\usepackage{graphicx}
\usepackage{tabularx, booktabs}
\usepackage{rotating}
\usepackage{hyperref}
\usepackage{esint} 
\usepackage{dsfont} 
\makeatletter
\AtBeginDocument{
  \hypersetup{
    linkcolor=blue,
    filecolor=magenta,      
    urlcolor=cyan,
    pdftitle={\@title},
    pdfauthor={Murdock G. Grewar},
  }
}
\makeatother
\usepackage{tikz}
\usetikzlibrary{positioning,spy,angles,quotes,calc}
\newcolumntype{R}{>{\raggedleft\arraybackslash}X}

\newcommand{\secref}[1]{\hyperref[#1]{\ref*{#1}\ \emph{\nameref*{#1}}}}

\newcommand{\diag}[1]{\operatorname{\mathrm{diag_{#1}}\!}}

\newcommand\zoominset[6]{ 
\begin{tikzpicture}[x=#1, y=#1, font=\footnotesize]
  \node[anchor=south west,inner sep=0] (image) at (0,0) {\includegraphics[width=#1]{#2}};
  \coordinate (viewport lower left)  at (#4);
  \coordinate (viewport upper right) at (#5);
  \draw[ultra thick, white] (viewport lower left) rectangle (viewport upper right);
  \coordinate (zoomwindowbottomleft) at (#6);
  \node[above right=0mm of zoomwindowbottomleft, draw=white, ultra thick, inner sep=0pt] (zoom1) {
       \scalebox{#3}{\tikz{
          \clip (#4) rectangle (#5);
          \node[anchor=south west,inner sep=0] (zoom2) at (0,0) {\includegraphics[width=#1]{#2}};
         }}%
       };
  \draw[ultra thick, white, dashed] (viewport lower left) -- (zoom1.south west);
  \draw[ultra thick, white, dashed] (viewport upper right) -- (zoom1.north east);
\end{tikzpicture}
}

\title{Practical Global Backprojection-Convolution in Transmission Cone-beam Computed Tomography}
 
\author{%
Murdock~G.~Grewar\thanks{M. Grewar is with the Department of Materials Physics, Australian National University, Australia.}
}
 
\IEEEpubid{0000--0000/00\$00.00~\copyright~2026 IEEE}

\begin{document}

\thispagestyle{empty}
\twocolumn[{%
© 2026 IEEE. Personal use of this material is permitted. Permission from IEEE must be obtained for all other uses, in any current or future media, including reprinting/republishing this material for advertising or promotional purposes, creating new collective works, for resale or redistribution to servers or lists, or reuse of any copyrighted component of this work in other works.%
}]
\newpage
\addtocounter{page}{-1}  

\maketitle

\begin{abstract}
Global backprojection-convolution (GBC) is a new framework for exact reconstruction in transmission cone-beam computed tomography (CBCT). GBC requires the X-ray source points to comprise a multidimensional `source locus' $X \subset \mathbb R^3$, such as a cylinder or sphere surrounding the object. GBC assumes that the attenuation coefficient is zero outside the reconstruction support (e.g. it is subject to the `long-object problem' for cylindrical scans). 
Theoretically, GBC is computationally highly expedient due to its structure, but producing a practical computational implementation poses a significant challenge because the method is uniquely vulnerable to four sources of discretisation error: (1) accurate discretisation of a multidimensional locus requires more points than for a 1-dimensional locus, (2) the convolution kernel has infinite range and so the backprojected volume must be of infinite size, (3) the discrete convolution kernel cannot be computed in closed form, and (4) aliasing artefacts in the backprojection are enormously magnified by the convolution step. In this article, we propose an assortment of discretisation regularisations, at the level of the symbolic algorithm, that mitigate these discretisation errors. As a prototype, we deploy the concept on the case where $X$ is a cylinder. The resulting algorithm is evaluated through a series of reconstructions of the 3D FORBILD head phantom, and a reconstruction from an experimental rock scan dataset.
\end{abstract}


\section{Introduction}\label{sec:introduction}

\begin{figure}[!t]
\begin{center}
    \subfloat[]{\includegraphics[width=0.33\linewidth]{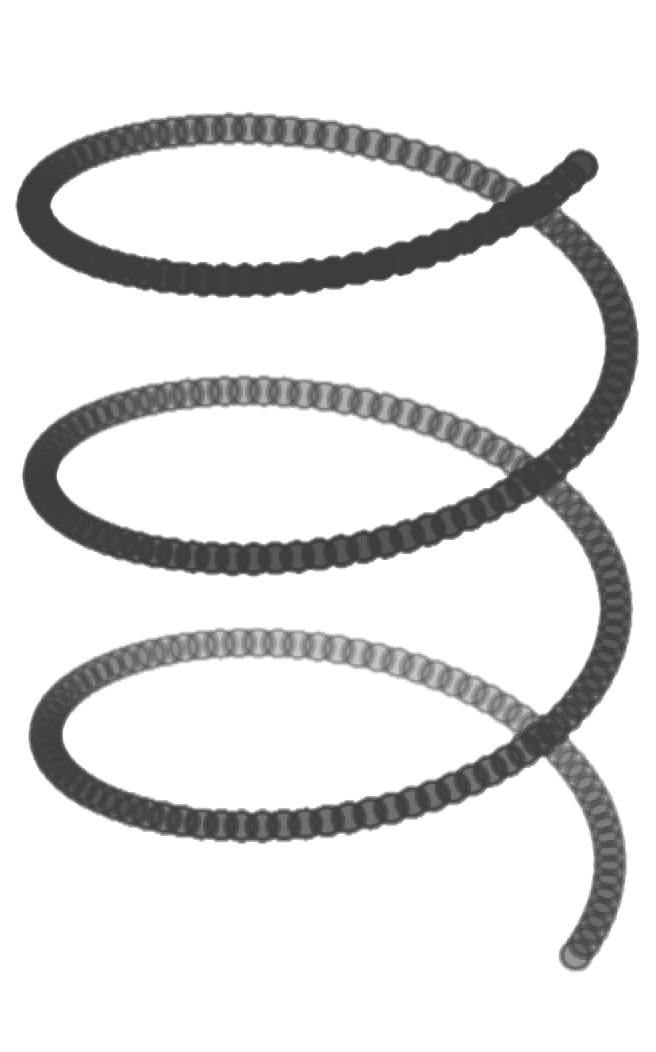}%
    \label{fig:trajectories_helix}}%
    \hfil
    \subfloat[]{\includegraphics[width=0.33\linewidth]{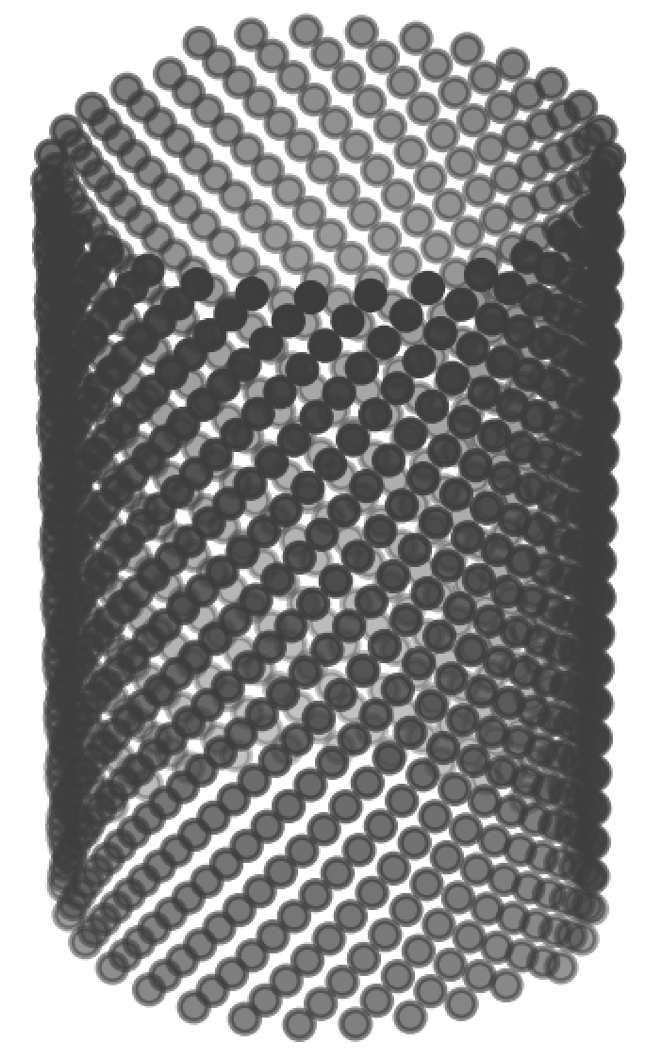}%
    \label{fig:trajectories_sft}}%
    \hfil
    \subfloat[]{\includegraphics[width=0.33\linewidth]{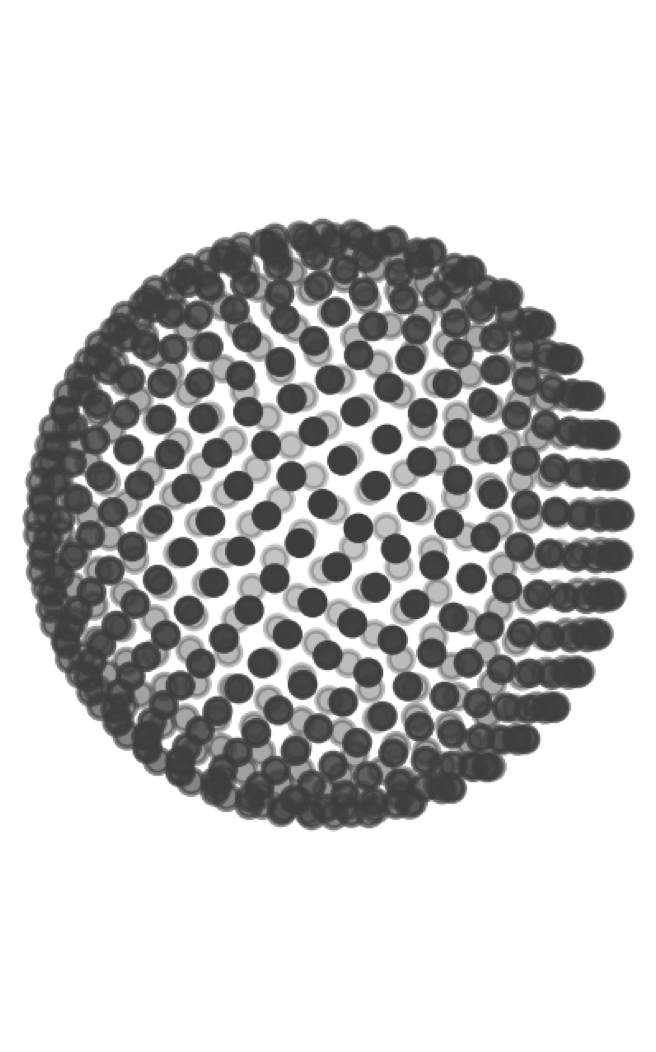}%
    \label{fig:trajectories_lds_sphere}}%
    \caption{Examples of acquisition loci in CBCT. The discrete points are the locations at which the X-ray source is placed as transmission measurements are made. The latter two loci are multidimensional and the associated measurement data are amenable to volume reconstruction via global backprojection-convolution (GBC).}
    \label{fig:trajectories}
\end{center}
\end{figure}

\IEEEPARstart{N}{umerous} methods have been proposed for direct inversion in transmission cone-beam computed tomography (CBCT) since 1995, e.g. \cite{tam1995, katsevich2002theoretically, katsevich2003general, ye2005general}, but most are limited to scanning loci that are 1-dimensional curves, such as a helix that wraps around the object (see Fig.~\ref{fig:trajectories_helix}). On the other hand, \emph{multi}dimensional scanning loci offer properties superior for materials imaging: they allow the positions of the cone-beam X-ray source to spread out and evenly occupy the space around the object, and this provides greater independence between the data, therefore greater statistical power, therefore fewer required projections, due to the wide coverage of viewing angles of the object \cite{kingston2018space}. See Figs.~\ref{fig:trajectories_sft} and \ref{fig:trajectories_lds_sphere}.
There are other advantages: the spreading out of source points makes it easier to correct for unintended X-ray source motion during scanning \cite{kingston2018space,delgado2019pi} and has been shown to assist in reducing metal artefacts from imaging printed metal parts \cite{kingston2020techniques}. It also alleviates the uneven resolution within tomograms that has been shown to occur in helix-like source loci \cite{varslot2012considerations}. 
There has been commercial interest in multidimensional scanning loci, with Siemens investigating the use of spherical loci \cite{bauer2021} and ``CT Lab'' at the Australian National University regularly conducting scans with cylindrical loci for commercial clients. 

Besides the merits established in these prior publications, multidimensional loci may also be motivated by future technology for CBCT imaging--- in particular, experimental setups involving multiple X-ray sources based on carbon nanotube technology \cite{patent_multisource_varex, patent_cnt_highcurrent_microx}. For example, a cylindrical locus may be realised by linearly translating a ring of sources.


\IEEEpubidadjcol 

Until recently, there had been no published method of direct reconstruction from multidimensional source loci. Either iterative methods have been employed---such as \cite{myers2016rapidly} which has been used in conjunction with the trajectory in \cite{kingston2018space} (see Fig.~\ref{fig:trajectories_sft}) at CT Lab---or the cone-beam data has had to be extrapolated and reformatted into a parallel-beam or Radon format in a process commonly referred to as `rebinning' (e.g. \cite{noo1997}), or a direct-but-approximate method was applied (e.g. \cite{barrett1994cone}).
In this article, we present a method for exact/direct inversion in CBCT that can be applied to a broad class of multidimensional loci. We refer to the resulting reconstruction technique as \emph{global backprojection-convolution} (GBC) because it is comprised of (1) a weighted `backprojection' of all measurement data, followed by (2) a global convolution of the backprojected volume. GBC may be considered a generalisation to CBCT of so-called `deconvolution algorithms' that apply to parallel-beam data and which were published in the Positron Emission Tomography (PET) literature in the 80s--90s \cite{colsher1980fully,schorr1983general,defrise1989three,defrise1995}. GBC cannot be applied to 1-dimensional source loci such as the helix.

\IEEEpubidadjcol

GBC is best understood as a system for constructing appropriate backprojection weightings which render the point-spread function of the imaging system translation-invariant. This enables the reconstructed volume to be produced as a volume deconvolution of the weighted cone-beam `backprojection' of the measurement data. A similar strategy is employed in PET imaging, in a technique known as `filtered backprojection' \cite{defrise2005image} or `FBP'. However, because in PET imaging the data is first rearranged into a parallel-beam format, it is more convenient there to apply the deconvolution to the measurement data, instead of the backprojected volume (this `commuting' of the filter with the backprojection is possible because the filter is `factorisable' \cite{defrise1989three}). GBC is a generalisation of these PET `deconvolution algorithms' \cite{schorr1983general,defrise1989three} to cone-beam imaging. A weighting term in the cone-beam backprojection emulates a parallel-beam backprojection, after which a volume deconvolution recovers the volume. This is the operating principle of GBC. No rebinning of the data is required.

GBC is---apart from being the first direct method of reconstruction on a new variety of CBCT datasets---also a high-performance, highly-parallelisable algorithm.
In a typical CBCT experiment, the measurement data is collected on a 2D detector (e.g. a flat-panel CCD) by situating the X-ray source at a fixed location (opposite the detector) and measuring the transmission of X-rays from the source through to each of the pixels of the detector (see Fig.~\ref{fig:cyl_pic}). The data is collected as a sequence of 2D `radiographs', and is linearised (i.e. converted to line integrals of attenuation from source to pixel) on a pixel-wise basis. Therefore, in the data format output by a typical CBCT scanning apparatus, the data is ordered first (large stride in computer memory) by the location of the source point, and second (small stride in computer memory) by the index of the detector pixel.  
GBC provides a particularly efficient means of direct reconstruction because it requires only a single sequential read of the cone-beam projections. Furthermore, the backprojection step (which forms the overwhelming bulk of the computational burden) is fully parallelisable for implementation on GPUs. 
The backprojection step may even take place concurrently with data acquisition. 

Due to the unbounded 3D deconvolution, several severe discretisation errors arise as one tries na\"ively to apply the continuum GBC result to discrete data. These errors, if left uncorrected, are significant enough to render GBC reconstruction impractical. Therefore, our goal in this article is to lay out a first practical implementation of GBC which quells the most significant discretisation errors. 

The outline of this paper is as follows. We begin in \S\ref{sec:inversion_theory} with a summary of the theory of GBC. Then, in \S\ref{sec:practical_discrete_methods}, we illustrate that the bare GBC algorithm suffers from severe discretisation artefacts. We then develop a symbolic algorithm to apply GBC more effectively to real, discrete data. This involves the introduction of four \emph{discretisation regularisations} which are designed to preserve the exactness of the inversion formula in the continuum limit. As our prototype, we apply GBC to the cylindrical locus, though the ideas can be generalised to other multidimensional loci. We present reconstructions from simulated and experimental data in \S\ref{sec:demonstrations}. The simulation demonstrates the effect of source point density on the quality of the reconstruction. The experimental reconstruction exhibits enhanced sharpness compared with the slower, iterated method in \cite{myers2016rapidly}, and greater resolution isotropy compared with a Katsevich \cite{katsevich2003general} reconstruction from double-helix data. 

\section{\texorpdfstring{Background:\\Global Backprojection-Convolution}{Background: Global Backprojection-Convolution}}\label{sec:inversion_theory}
This section summarises the CBCT inversion problem and the main ideas behind the theory of GBC. 

\subsection{The Inversion Problem}
The inversion problem in transmission X-ray computed tomography (XCT) is to reconstruct a 3-dimensional greyscale image---called a \emph{tomogram}---of an object that has been scanned by penetrating X-rays. The problem is modelled mathematically by supposing that we have experimentally determined the numeric values for a set of line integrals of the ideal tomogram $\mathbf x$. These measurements of the line integrals are arranged in a vector, which we call $\mathbf m$, and the basic mathematical model of the imaging process is given by the Beer-Lambert law \cite{mayerhofer2020bouguer} as the linear equation
\begin{equation}
    \mathbf m = M \mathbf x \, .
\end{equation}
The operator $M$ is a linear integral operator that integrates the attenuation coefficient in $\mathbf x$ along line segments from the X-ray source point to the detector pixels, at the various combinations of source and detector positions.
$M$ is known from the geometry of the X-ray experiment. The basic goal of XCT is to produce a computer representation of $\mathbf x$, given the measurements $\mathbf m$ and the geometry of the experiment.
Therefore, the theoretical side of XCT research essentially concerns the construction of practical computational algorithms that implement a \emph{left inverse} of $M$, for various geometries. A left inverse of $M$ is any operator $M^{-L}$ that satisfies $M^{-L} M = \mathrm{Id}$. Applying a left inverse to the measurement data solves the inversion problem viz. $\mathbf x = M^{-L} \mathbf m $. 
We will use the Moore-Penrose left inverse in this article.

In reality, there are various complications to this simple picture which arise from imperfections in the imaging process and the validity/applicability of the Beer-Lambert law. 
Such complications are out-of-scope for this article, and are often ignored in transmission CT. One example is `beam-hardening' of polychromatic beams, resulting in beam-hardening artefacts in reconstructed tomograms \cite{brooks1976beam}.

\subsection{Theory of Global Backprojection-Convolution (GBC)}
\label{sec:inversion_theory_background}
The continuum theory of inversion via global backprojection-convolution (GBC) extends to CBCT a series of results on `deconvolution algorithms' that were established for Positron Emission Tomography (PET), at first by Colsher in a special case \cite{colsher1980fully}, then by Schorr more generally \cite{schorr1983general}, then extended by Defrise \emph{et al.} \cite{defrise1989three,kinahan1989analytic,defrise1995}. This body of theory differs substantially from the Radon-transform-based body of theory (esp. \cite{grangeat1991}) that underpins many existing methods of inversion in CBCT. GBC is suitable for use on scanning loci that are \emph{multi}dimensional, but not for one-dimensional trajectories such as the helix; see Fig.~\ref{fig:trajectories}.

\subsubsection{Parallel-beam inversion via GBC}\label{sec:gbc_parallel_theory}
We now summarise how GBC may be applied to parallel-beam data. This is necessary because the cone-beam theory builds on the parallel-beam theory. The particular results relating to the cylindrical trajectory will be procured as a special case in \S\ref{sec:specialisation_to_cylinder}. The parallel-beam theory was established previously in \cite{defrise1989three}; we provide here a summary in a format which is more convenient for its extension to cone-beam data.

Let $\mathcal A = \mathcal S(\mathbb R^3, \mathbb R)$ be the Schwartz space of attenuating volumes (i.e. suitably `well-behaved' tomograms, which are real-valued functions on $\mathbb R^3$). Let $S^2$ be the sphere of unit vectors. Let $\mathcal M = \mathcal S(T(\mathbb RP^2), \mathbb R)$ be the Schwartz space of real-valued functions on the tangent bundle of the real projective plane $\mathbb RP^2$. A `measurement field' $m \in \mathcal M$ assigns a real number to every (nondirected) line in $\mathbb R^3$. Notationally, a line will be specified by its direction $\hat \theta \in S^2$ and by a point $r \in \mathbb R^3$ included in the line. For a measurement field $\mathbf m \in \mathcal M$, we denote the value assigned to the respective line by $m(\hat \theta, r)$. Let $A : \mathcal A \rightarrow \mathcal M$ be the X-ray transform, which maps attenuating volumes to their line projections (cf. \cite[\S{}II.1]{natterer2001mathematics}). 

Suppose that a complete set of experimental measurements is taken along every measurement line in 3-dimensional space, producing a measurement field $\mathbf m \in \mathcal M$.
Then, for every choice of $L^2$ inner product on $\mathcal M$, there is a tomogram $\mathbf x \in \mathcal A$, unique up to $L^2$ differences, whose X-ray transform $A \mathbf x$ minimises the normed difference between $A \mathbf x$ and $\mathbf m$.
It is produced by applying the Moore-Penrose left inverse $A^{+} = (A^{\dag} A)^{-1} A^{\dag}$ to $\mathbf m$:
\begin{equation}\label{eq:parallel_gbc}
    \mathbf x
    =
    \underset{\mathbf x' \in \mathcal A}{\mathrm{argmin}} || \mathbf m - A \mathbf x'||^2
    =
    (A^{\dag} A)^{-1}
    \underbrace{A^{\dag}}_{{\text{backprojection}}} \; 
    \underbrace{\mathbf m}_{\text{measurements}} \, .
\end{equation}
The norm $||\cdot||$ and the `backprojection' $A^\dag$ (and therefore $A^+$) are dependent on the choice of inner product on $\mathcal M$. The reason that the GBC algorithm is possible is because the operator $(A^\dag A)^{-1}$ happens to be a volume convolution for certain choices of inner product. This is a desirable form because convolutions are relatively cheap to compute, meaning that we can compute the optimal solution $\mathbf x$ in an efficient manner.

We will consider inner products on $\mathcal M$ of the following form, where the function $f : S^2 \rightarrow \mathbb R^+$ may be freely chosen. The symbol $\hat \theta^\perp$ denotes the 2D subspace of $\mathbb R^3$ orthogonal to $\hat \theta$:
\begin{equation}\label{eq:inner_product_meas}
    \langle \mathbf m_1, \mathbf m_2 \rangle_f = \iint_{\mathrlap{S^2}} \; \mathrm d^2 \hat \theta f(\hat \theta) \iint_{
        \mathrlap{
            \hat \theta^\perp
        }} \mathrm d^2 r \; m_1(\hat \theta, r) m_2(\hat \theta, r) \, .
\end{equation}
This is the most general form for the inner products on $\mathcal M$ which result in a GBC algorithm.
With this choice of inner product, the operator $(A^\dag A)^{-1}$ is the following convolution operator. Let $F_{\mathcal A}$ be the 3-dimensional Fourier transform operator acting on attenuating volumes in $\mathcal A$. Let $\xi \in \mathbb R^3$ represent the 3D frequency vector (cycles per unit length). The operator $\mathcal I$ will be described subsequently. The convolution operator is 
\begin{equation}\label{eq:deconv_filter_int}
    (A^{\dag} A)^{-1} 
    = F_{\mathcal A}^{-1} \diag{\xi}\left( \frac{|\xi|}{\mathcal I[f](\hat{\xi})} \right) F_{\mathcal A}\, ,
\end{equation}
where $\diag{\xi}(\dots)$ is the diagonal operator that multiplies the spatial frequency indexed by $\xi$ by the formula $(\dots)$ involving $\xi$. We denote $\hat{\xi} = \xi / |\xi|$. 
The operator $\mathcal I$, known as the Funk transform \cite{funk1911flachen}, transforms functions on $S^2$ like so:
\begin{equation}\label{eq:deconv_integral_expression}
\mathcal I[f](\hat{\xi}) 
= \iint_{S^2} \mathrm d \hat \theta \delta(\hat \theta \cdot \hat{\xi}) f(\hat \theta)
= \int_{\{\hat \theta \in S^2 : \hat \theta \cdot \hat \xi = 0\}} \mathrm d \hat \theta f(\hat \theta) \, ,
\end{equation}
where `$\cdot$' denotes the standard vector dot product on $\mathbb R^3$, and `$\delta$' is the Dirac-delta distribution. An illustrated example of the Funk transform will be provided later in Figure~\ref{fig:orlov}. An appropriate choice of $f$ will be constructed for arbitrary geometries in \S\ref{sec:inversion_theory_background}.

The backprojection operator $A^{\dag}$ is the adjoint of the X-ray transform $A$, with respect to the chosen inner product, and thus depends on $f$. It acts on $\mathbf m$ to produce a new attenuating volume whose attenuation coefficient at $r \in \mathbb R^3$ is
\begin{equation}\label{eq:backproj_parallel}
    (A^{\dag} \mathbf m)(r) = \iint_{\mathrlap{S^2}} \; \mathrm d^2 \hat \theta f(\hat \theta) m(\hat \theta, r) \, .
\end{equation}
We stipulated above that $f$ be a \emph{positive}-valued function on $S^2$. However, these results extend to some \emph{nonnegative} $f$. In particular, the inverse \eqref{eq:deconv_filter_int} exists (and the tomogram minimising the norm \eqref{eq:inner_product_meas} is also unique) for $f$ which satisfy\footnote{
    Strictly speaking, a stronger condition than $\mathcal I[f](\hat \theta) \neq 0$ may be required, e.g. that $\inf_{\hat \theta \in S^2}|\mathcal I[f](\hat \theta)|$ be nonzero. 
}: $\forall \hat \theta \in S^2, \mathcal I[f](\hat \theta) \neq 0$. In \cite{defrise1989three}, this is referred to as `Orlov's condition'. 
The backprojection \eqref{eq:backproj_parallel} uses measurement data from projection lines parallel to $\hat \theta \in S^2$ only where $f(\hat \theta) \neq 0$. Therefore, \emph{GBC does not actually require measurements along all line directions}; $f$ may be chosen to restrict the required line directions.
For a parallel-beam experiment modelled by a continuous distribution of beam directions on $S^2$, a natural choice for $f$ is that distribution of beam directions. 

\subsubsection{Cone-beam inversion via GBC}\label{sec:conebeam_inversion_derivation}
We now describe how GBC is applied to cone-beam data. The main theoretical innovation of this article is to observe that for suitable multidimensional scanning loci, a weighting term may be inserted into the cone-beam backprojection so that it emulates a parallel-beam backprojection. 

The cone-beam problem is solved by drawing an equivalence with the parallel-beam problem. First, examine the parallel-beam GBC reconstruction equation \eqref{eq:parallel_gbc}. The first operation performed in that equation is to backproject the measurement data $\mathbf m$ into the volume, with the backprojection operator $A^\dag$. The second operation is a volume convolution. The cone-beam reconstruction equation is produced by implementing the parallel-beam backprojection $A^\dag$ as a cone-beam backprojection. For certain geometries, this can be acccomplished efficiently by adding a simple `weighting' term to the cone-beam backprojection. This weighting term can be computed as a closed-form expression for some geometries, and in this case, there is negligible computational overhead required in addition to a standard cone-beam backprojection. We will now elaborate on the details.

For the cone-beam backprojection, assume a distribution of source points $\chi(x)$ over $x \in \mathbb R^3$. Let $W(x,r)$ be a weighting term dependent on the source location $x$ and the point $r$ in the volume at which the backprojection is calculated. The basic cone-beam backprojection from a single source point $x$ is denoted $B_x$ and is expressed by the equation
\begin{equation}
	(B_x \mathbf m)(r) = m \left(\frac{r - x}{|r - x|} , r \right) \, .
\end{equation}
The weighted cone-beam backprojection, call it $B$, is a weighted sum over the distribution of source points. It is given by
\begin{align}
	B &= \iiint_{\mathrlap{\mathbb R^3}} \; \mathrm d^3 x \; \chi(x) W_x B_x \\
	(B \mathbf m)(r) &= \iiint_{\mathrlap{\mathbb R^3}} \; \mathrm d^3 x \; \chi(x) W(x,r) m \left(\frac{r - x}{|r - x|} , r \right) \, .
\end{align}
Note that when a discrete point distribution is used for $\chi$, as in a real experiment, then the triple-integral amounts to a sum over the X-ray source points $x_i$ viz.
\begin{align}
	\chi(x) &= \sum_i \delta^3(x-x_i) \, ,\\
	\Rightarrow \quad (B \mathbf m)(r) &= \sum_{i} W(x_i,r) m \left(\frac{r - x_i}{|r - x_i|} , r \right) \, .
\end{align}

For source point distributions $\chi$ which (loosely speaking) `surround' the object support, it becomes possible to choose the weighting function $W(x,r)$ in such a way that the cone-beam backprojection $B \mathbf m$ is identical to the parallel-beam backprojection $A^\dag \mathbf m$, for some choice of $f$. Therefore, with an appropriate choice of $\chi$, $f$ and $W(x,r)$, the reconstruction equation for cone-beam tomography reduces to the Global Backprojection-Convolution (GBC) equation
\begin{align}\label{eq:cone_gbc}
	\mathbf x = \underset{\mathbf x' \in \mathcal A}{\mathrm{argmin}} ||\mathbf m - A \mathbf x'||^2
	= \underbrace{(A^\dag A)^{-1}}_{\mathclap{\text{volume deconvolution}}} \quad \overbrace{B \mathbf m}^{\mathclap{\text{cone-beam backprojection}}} \, .
\end{align}
This is equivalent to the parallel-beam reconstruction in Equation~\eqref{eq:parallel_gbc}, except with the parallel-beam backprojection $A^\dag \mathbf m$ replaced by the weighted cone-beam backprojection $B \mathbf m$. This makes for a highly efficient reconstruction algorithm from cone-beam measurement data.

The conditions under which exact reconstruction via GBC is possible are as follows. Let $V \subset \mathbb R^3$ be the reconstruction support. Assume that the object's attenuation coefficient is $0$ outside $V$. (Note that this requirement introduces the `long-object problem'.) Define the function $d(v; \hat \theta)$ as the \emph{solid-angular density of source points seen from $v$ in the direction of $\hat \theta$}--- this depends on $\chi$. A source point $x$ is said to be `seen' from $v$ if the experiment captures a measurement line from source point $x$ which includes the point $v$. Let $\Omega \subseteq S^2$ be the \emph{common antipodal Orlov sphere support over $V$}, defined as
\begin{equation}\label{eq:simp_weight_C}
    \Omega = \left\lbrace \! \hat \theta \in S^2 \!:\! \inf_{v \in V} \left\lbrace
        d(v; \hat \theta) \right\rbrace \!>\! 0 \text{ and } \inf_{v \in V} \left\lbrace d(v; -\hat \theta) 
    \right\rbrace \!>\! 0 \right\rbrace .
\end{equation}
Geometrically, $\Omega$ is the region of $S^2$ over which all $v \in V$ see source points in all directions contained in $\Omega$ and in the opposite directions.\footnote{We have required that source points be seen in \emph{both} directions because otherwise the backprojected volume will contain singularities at the locations of source points. This poses issues for representing the backprojection on a voxelised lattice.} Exact reconstruction over $V$ is possible when the Funk transform of $\Omega$ is everywhere nonzero.\footnote{We have used almost the same definition for $\Omega$ as in \cite{defrise2005image} which regards parallel-beam imaging in PET. The masking of projections to the directions contained in $\Omega$ is a generalisation of the masking that Colsher proprosed in \cite{colsher1980fully}. This masking is necessary to ensure that the cone-beam forward-backprojection has a translation-invariant point-spread; this makes it equivalent to a parallel-beam geometry, and hence the volume may be recovered via a deconvolution.} 

When GBC reconstruction is possible, then a sufficient weighting function $W$ is given by
\begin{equation}\label{eq:simp_weight_01}
   W(x, r) = \begin{cases}
	   \frac{1}{\frac 1 2 \left(d(r; x-r) + d(r; r-x)\right)} \text{ if } \frac{x-r}{|x-r|} \in \Omega \\
       0 \text{ otherwise.}
   \end{cases} \, ,
\end{equation}
which corresponds to the following choice of $f$:
\begin{align}\label{eq:GBC_f}
	f(\hat \theta) &= \begin{cases}
		1 \text{ if } \hat \theta \in \Omega \\
		0 \text{ otherwise.}
	\end{cases} \, .
\end{align}
The corresponding GBC equation for reconstruction is
\begin{align}\label{eq:cone_beam_inversion_formula}
    \mathbf x 
    \!&=\! \underbrace{F_{\mathcal A}^{-1} \mathrm{diag}_{\xi}\Bigg( 
        \frac{|\xi|}{\mathcal I[f](\hat{\xi})}
    \Bigg) F_{\mathcal A}}_{\text{volume deconvolution}} \underbrace{\left( \iiint_{\mathbb R^3} \mathrm d x \chi(x) W_x B_x \right)}_{\text{weighted backprojection}} \mathbf m \, .
\end{align}
Closed-form expressions can be produced for both $W(x,r)$ and $\mathcal I[f](\hat{\xi})$ for many choices of the source point distribution $\chi(x)$. We will not elaborate on this matter except to point out that the cylindrical and spherical source point distributions (e.g. see Fig.\ref{fig:trajectories}) yield closed-form expressions.

Equivalency between the weighted cone-beam backprojection $B$ and the parallel-beam backprojection $A^\dag$ may be established as follows:
\begin{align}
    &(B\mathbf m)(r) = \iiint_{\mathrlap{\mathbb R^3}} \mathrm d^3 x \; \chi(x) W(x,r) m\left(\frac{r-x}{|r-x|} ,r\right) \nonumber \\
    &= \iint_{\mathrlap{\Omega}} \mathrm d^2 \hat \theta \; \frac{d(r; \hat \theta)}{\tfrac{1}{2} (d(r; \hat \theta) + d(r; -\hat \theta))} m(\hat \theta, r) \tag{change of coordinates; using Eq.~\ref{eq:simp_weight_01} for $W(x,r)$} \nonumber \\
    &= \iint_{\mathrlap{\Omega}} \mathrm d^2 \hat \theta \; \frac{\tfrac{1}{2}(d(r; \hat \theta) + d(r; -\hat \theta))}{\tfrac{1}{2} (d(r; \hat \theta) + d(r; -\hat \theta))} m(\hat \theta, r) \tag{antipodal symmetrisation using $\hat \theta \in \Omega \Rightarrow -\hat \theta \in \Omega$} \nonumber \\
    &= \iint_{\mathrlap{\Omega}} \mathrm d^2 \hat \theta \; m(\hat \theta, r) \nonumber \\
    &= \iint_{\mathrlap{S^2}} \mathrm d^2 \hat \theta \; f(\hat \theta) m(\hat \theta, r) \text{ with $f$ given in Eq.~\ref{eq:GBC_f} } \nonumber \\
    &= (A^\dag \mathbf m)(r) \text{ with $f$ given in Eq.~\ref{eq:GBC_f}; $A^\dag$ given in Eq.~\ref{eq:backproj_parallel}} \, .
\end{align}

\subsection{Specialisation to the Cylindrical Locus}\label{sec:specialisation_to_cylinder}

The results established in \S\ref{sec:inversion_theory_background} apply to any CBCT imaging geometry. In this section, we specialise those results to the cylindrical source trajectory. 
The cylindrical source locus is of singularly broad appeal because cylindrical scans can already be performed using most existing CBCT apparatuses, e.g. medical industrial CT scanners and research laboratory scanners. A good choice of cylindrical locus is found in \cite{kingston2018space} (see Fig.~\ref{fig:trajectories_sft}), but any locus may be used provided that it uniformly samples the cylinder (i.e. equal number of source points per unit area). Note that GBC with the cylindrical locus does suffer from the so-called `long-object problem' as defined in \cite{danielsson97:proceedings_3D_1997}. Essentially, this means that we need to assume that the whole object is scanned (the projections are permitted to be axially truncated). Practically speaking, however, we will find in \S\ref{sec:experimental_recon} that this requirement may be relaxed.

\begin{figure}[!t]
\begin{center}
    \subfloat[]{\includegraphics[width=0.38\linewidth]{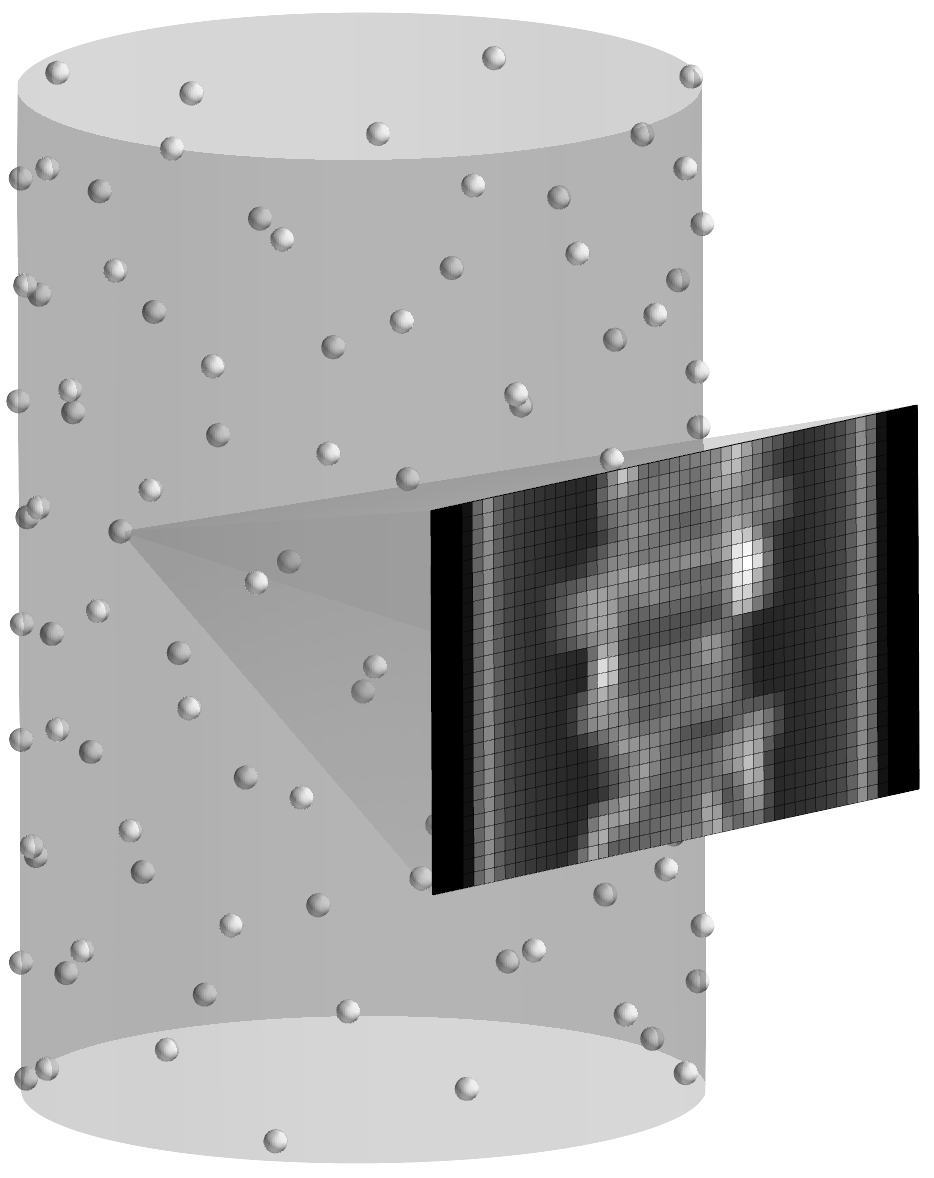}%
    \label{fig:cyl_pic}}%
    \hfil%
    \subfloat[]{
    \resizebox{0.6\columnwidth}{!}{%
    \begin{tikzpicture}[
        my angle/.style={draw, <->, angle eccentricity=1.3, angle radius=9mm},
        extended line/.style={shorten >=-#1,shorten <=-#1},
        extended line/.default=0.5cm]
        
        \node[circle,
           draw=lightgray,
           text=white,
           fill=lightgray,
           minimum size=2*1.0cm] () at (0,0){};

        \coordinate[circle, inner sep=0.5pt]                     (O)  at (0,0);
        \coordinate[circle, fill, inner sep=1.5pt, pin= 0:{$x$}]  (S)  at (2,0);
        \coordinate[] (e) at (140:2);
        \coordinate[circle, fill, inner sep={1.5pt}, pin= 200:{$v$}] (p) at ($(S)!.8!(e)$);
        \coordinate[] (c) at ($(S)!.5!(e)$);
        \coordinate[gray, circle, fill, inner sep={1.5pt}, pin= 55:{$y$}] (D) at ($(S)!1.2!(e)$);
        \coordinate[]  (L)  at (-2,0);

        \draw[transform canvas={xshift=0mm, yshift=0mm}, <->] (S) -- (O) node[midway, below]{$R$};
        \draw[transform canvas={xshift=1mm, yshift=1mm}, <->] (S) -- (p) node[midway, above right]{$\ell$};
        \draw[<->] (O) -- (p) node[midway, left]{$\rho$};
        \draw[transform canvas={xshift=0mm, yshift=0mm}, <->] (O) -- (230:1cm) node[midway, below right]{$R_V$};

        \draw pic["$\theta_h$",draw=black,<->,angle eccentricity=1.2,angle radius=1.5cm] {angle=p--S--O};

        \draw[thick, gray] let \p1=(D) in (\x1,-2.1) node[below, red] {} -- (\x1,2.1);
        \draw[gray] let \p1=(D) in (\x1,0.3) node[above, rotate=90] {detector};

        \draw ([shift=(-20:2cm)] 0,0) arc (-20:340:2cm);
        \draw[thick, dashed] (S) -- (D);
    \end{tikzpicture}
    }
    \label{fig:circle_diagram}}
\end{center}
\caption{An illustration of the imaging geometry with a cylindrical source locus. \textbf{(a):} The cylindrical locus is approximated by a discrete collection of source points. For our purposes, the X-ray detector is rectangular, and always oriented opposite the X-ray source in a natural way, and is horizontally and vertically centred with the source point it opposes.
We assume a uniform density of points on the surface of the cylinder. The precise distribution of points has no effect on the inversion theory or the inversion algorithm, although it can affect the reconstruction in practice. 
\textbf{(b):} A diagram of geometric quantities $\ell, \rho$ and $\theta_h$. 
The cylinder axis goes into the page. The point $x$ is the location of an X-ray source point on the cylinder, and $v$ is the location of a volume point into which projection data from $x$ is to be backprojected. The distances $\ell$ and $\rho$ and the angle $\theta_h$ are all measured in the 2D projection, as depicted. The measurement at $y$ is the integrated attenuation coefficient along the line $\vec{xy}$. The radius of the reconstruction support is $R_V$.
}
\end{figure}

Figure~\ref{fig:cyl_pic} contains an illustration of the imaging geometry.  
We denote the radius of the cylinder by $R$, the width of the detector by $W$, the height of the detector by $H$, and the source-to-detector distance by $L$. We assume that the cylinder has a uniform density of source points per unit area, $\mu$. We define the following two angles:
\begin{align}
	\vartheta_h &= \arctan(W/(2L)) \, , \\
	\vartheta_v &= \arctan \left( H/\sqrt{W^2 + 4L^2} \right) \, .
\end{align}
The angle $\vartheta_h$ is half the angle subtended by the middle row of the detector, from the X-ray source. The angle $\vartheta_v$ is half the angle subtended by the left (or right) column of the detector, from the X-ray source.

The goal of this subsection is to specialise the GBC reconstruction equation, i.e. Equation~\eqref{eq:cone_beam_inversion_formula}, to this imaging geometry. This requires us to choose an appropriate reconstruction support $V$, then solve for: the \emph{common antipodal Orlov sphere support} $\Omega$; an expression for the Funk transform of $f: S^2 \rightarrow \mathbb R$, where $f = 1$ on $\Omega$ and $0$ otherwise; and finally, the corresponding weighting function $W(x,r)$.

We begin by identifying an appropriate reconstruction support $V$. It is easiest to start with the assumption that the source point cylinder is infinitely tall, and later reduce the size of $V$ to account for the finite height of the source point cylinder. The maximal reconstruction support $V$ can be determined by applying the Smith criterion \cite{smith1985}, or another similar data sufficiency criterion. The maximal support $V$ consists of those points $v \in \mathbb R^3$ for which any plane containing $v$ also contains a source point `seen' by $v$. Assuming that the source points form a continuous surface, then $V$ turns out to be the cylinder coaxial with the source cylinder, but with reduced radius $R_V = R \sin \vartheta_h$. The reduced radius corresponds to the requirement that the projection data must not be transaxially truncated.

\begin{figure}[!t]
\begin{center}
	\subfloat[]{\includegraphics[height=3cm]{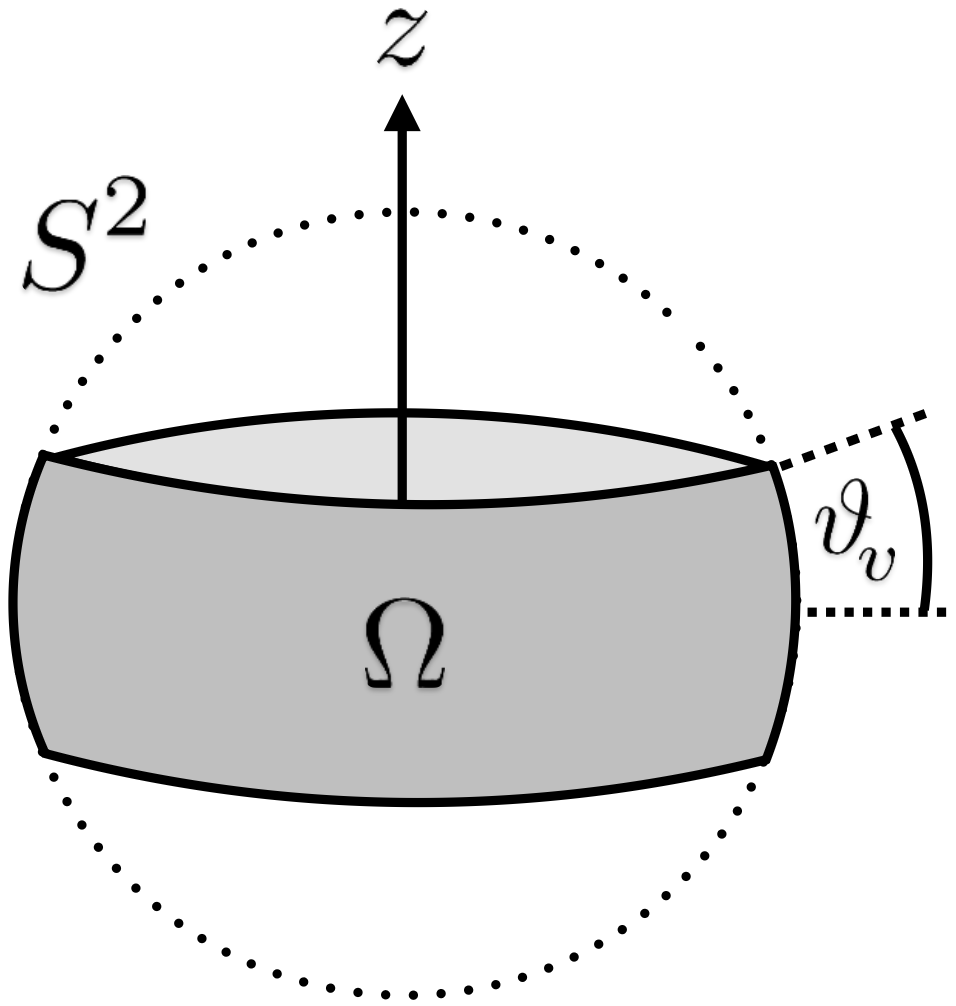}%
	\label{fig:orlov_a}}
	\hfil
	\subfloat[]{\includegraphics[height=3cm]{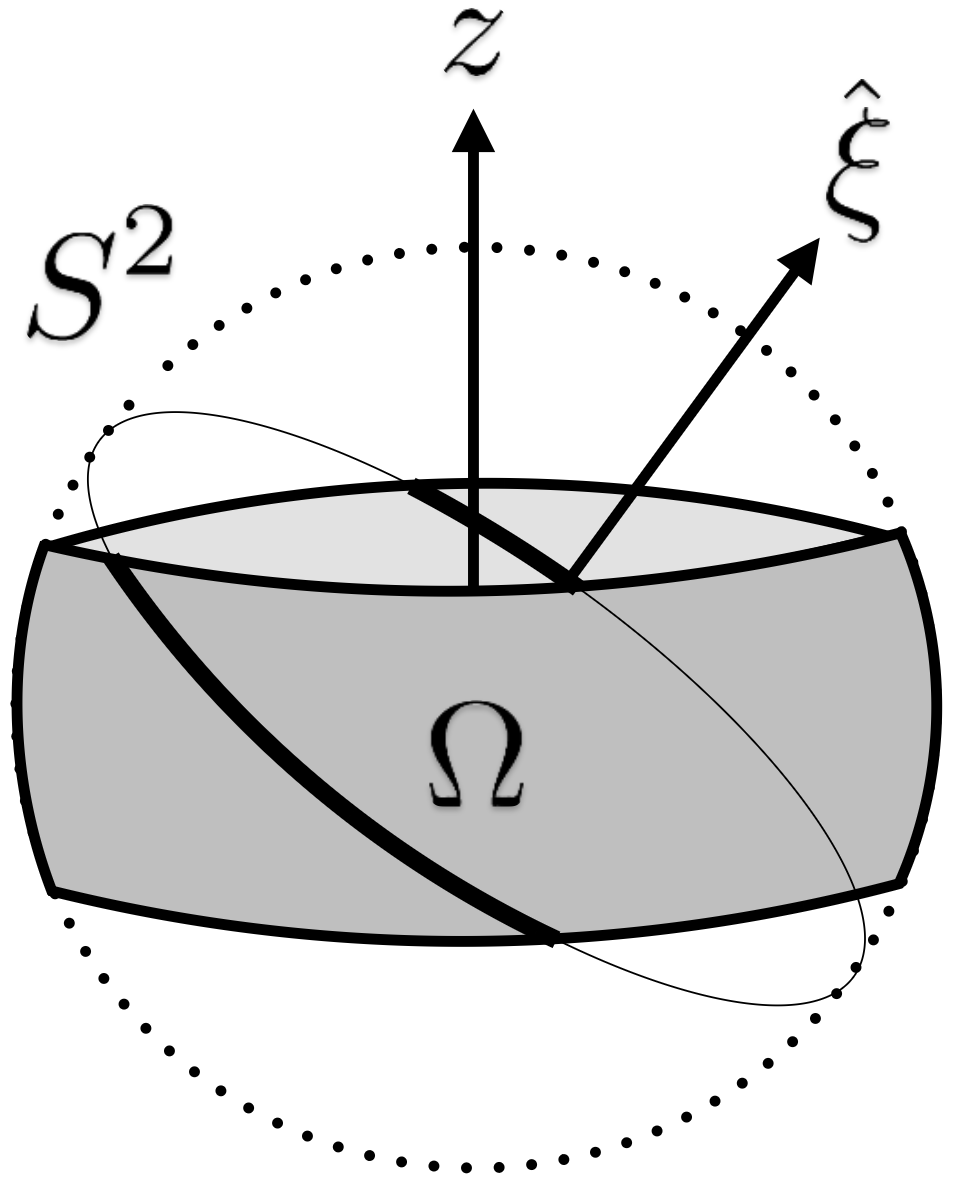}%
	\label{fig:orlov_b}}
\end{center}
\caption{A depction of the \emph{common Orlov sphere support} $\Omega$ for the cylindrical source locus with a rectangular detector. \textbf{(a):} $\Omega$ is seen to be an `equatorial band' with angular span $2 \vartheta_v$. \textbf{(b):} the Funk transform of $f$ evaluated at $\hat{\xi}$, i.e. $\mathcal I[f](\hat{\xi})$, is the length of the intersection of $\Omega$ with the great circle orthogonal to $\hat{\xi}$.}
\label{fig:orlov}
\end{figure}

Second, we solve for the \emph{common Orlov sphere support} $\Omega \subseteq S^2$, using the expression in Equation~\eqref{eq:simp_weight_C}. In spherical polar coordinates $(\theta, \phi)$ (with $\theta \in [0,\pi]$ the polar coordinate, and $\phi \in [0,2\pi)$ the azimuthal), it is given by 
\begin{equation}\label{eq:cylspec_C}
    \Omega = \left\lbrace \big.(\theta, \phi) : |\theta - \tfrac \pi 2| < \vartheta_v \right\rbrace \, .
\end{equation}
This selects the same range of projection angles as Colsher's method \cite{colsher1980fully}. The set $\Omega$ is illustrated in Figure~\ref{fig:orlov_a}. Other measurement data will be discarded, as depicted later in Figure~\ref{fig:colsher_hardmask}. 

Third, we solve for the Funk transform of $f$. The geometric meaning of this transform is illustrated in Figure~\ref{fig:orlov_b}. The same transform is used in filtered backprojection reconstruction in PET imaging, e.g. see \cite{defrise2005image}, though the transform has not always been referred to by name in the literature. In fact, the Funk transform of $f$ in the present case is the same one depicted in \cite[Fig.~4.10]{defrise2005image}. It is equal to
\begin{equation}\label{eq:cyl_eqs_I}
    \mathcal I[f](\hat{\xi}) = 4 \arcsin \left( \left. 
		    \max \left\lbrace 1 , \; \frac{\sqrt{1-\hat{\xi}_z^2}}{\sin \vartheta_v} \right\rbrace
    \right.^{-1} \right) \, ,
\end{equation}
where $\hat{\xi} \in S^2$ and $\hat{\xi}_z$ is its $z$ component.

Fourth and finally, we need to compute the corresponding weighting function $W(x,r)$ as defined in Equation~\eqref{eq:simp_weight_01}. This will complete our derivation of the specialisation of the GBC reconstruction equation \eqref{eq:cone_beam_inversion_formula} to this imaging geometry. This calculation involves consideration of the solid-angular density of source points, $d(v; \hat \theta)$, which was introduced earlier. 
The calculation of $W(x,r)$ involves some trigonometry. The result is as follows. Let $r \in \mathbb R^3$ be given in cylindrical polar coordinates by $(r_z, r_\rho, r_\phi)$. Likewise, let the source point $x \in \mathbb R^3$ be given in cylindrical polar coordinates by $(x_z, R, x_\phi)$. We define two angles:
\begin{subequations}
\begin{align}
    \theta_{\text{el}} &= \arcsin \frac{r_z - x_z}{|r - x|} \, , \\
    \theta_{h} &= \arcsin \frac{r_\rho \sin(r_\phi - x_\phi)}{\sqrt{r_\rho^2 + R^2 - 2 r_\rho R \cos(r_\phi - x_\phi)}} \, .
\end{align}
\end{subequations}
The angle $\theta_{\text{el}}$ is the angle of elevation of the point $r$, as seen from the source point $x$. The angle $\theta_{h}$ is similar, except it measures the horizontal angle of deviation; this is depicted in Figure~\ref{fig:circle_diagram}. Note that both angles apply equally well to points $r$ outside of the cylinder of radius $R$, as well as those inside. Adding/subtracting $\pi$ from $\theta_h$ will not alter the equation for the weighting. The equation for the weighting is
\begin{equation}\label{eq:cyl_eqs_weights}
    W(x,r) \!=\! \begin{cases}
	\frac{1}{\mu R^2} (\cos \theta_{\text{el}})^3 \frac{|\cos \theta_h|}{\cos(2 \theta_h) + (r_\rho/R)^2} \text{ if } |\theta_{\text{el}}| < \vartheta_v \\
    0 \text{ otherwise.}
\end{cases}
\end{equation}

Equation~\eqref{eq:cone_beam_inversion_formula} provides the reconstruction formula in terms of the Funk tranfsorm $\mathcal I[f]$ and the weighting function $W(x,r)$, which are provided in Equation~\eqref{eq:cyl_eqs_I} and Equation~\eqref{eq:cyl_eqs_weights} respectively. This completes our review of the cylindrical GBC formula in the continuum.

\section{Discretisation Regularisations}
\label{sec:practical_discrete_methods}
We have summarised the continuum theory of global backprojection-convolution (GBC), and the special case of a cylindrical scanning locus. Further work is required to develop the continuum result into a practical algorithm that works on discrete datasets, and that is the subject of this article.

\begin{figure}
    \begin{centering}
    \includegraphics[width=\linewidth]{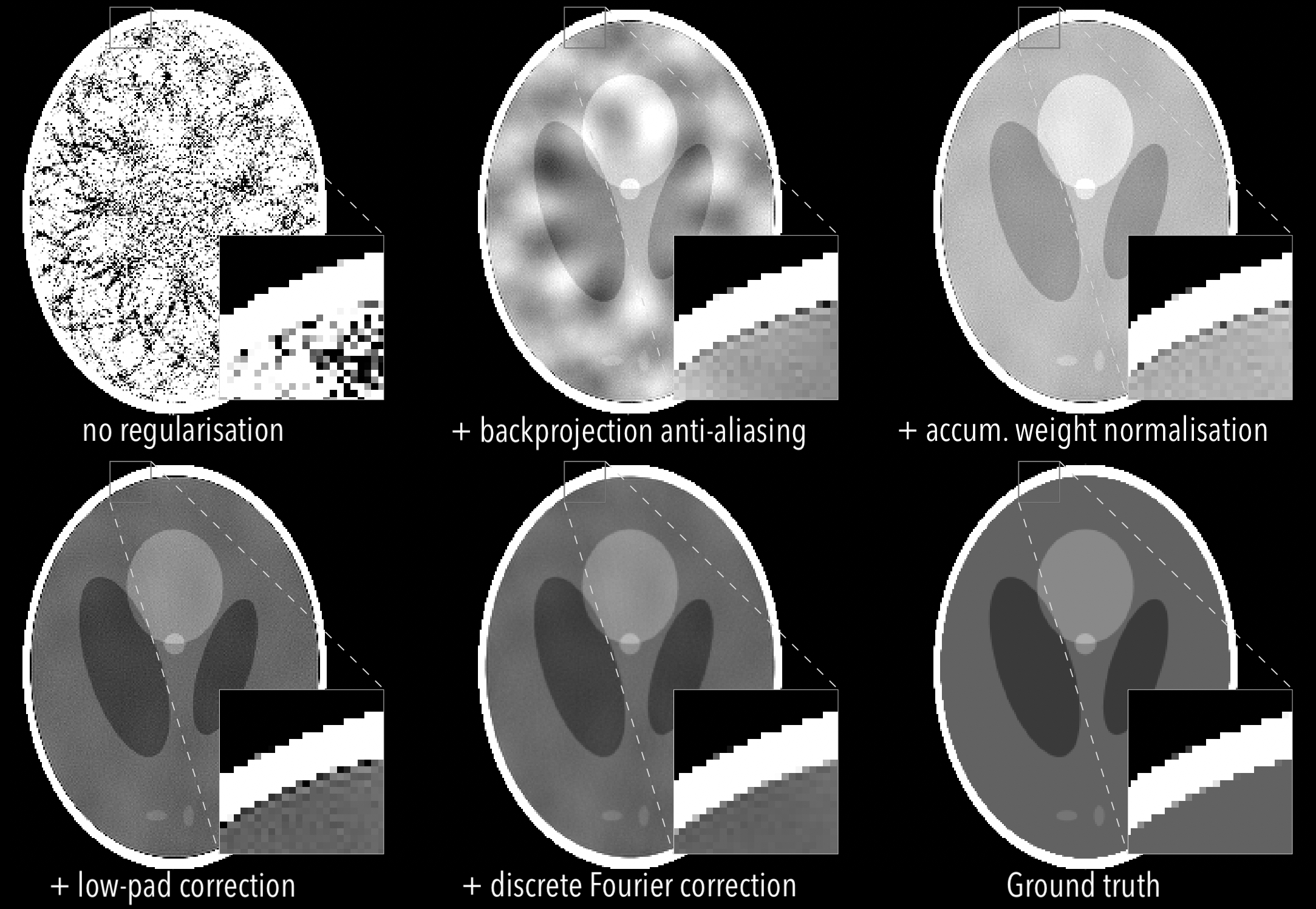}
    \caption{An illustration of the effects of adding the discretisation regularisations described in \S\ref{sec:practical_discrete_methods}, using a common 2D slice from reconstructions of the 3D Shepp-Logan phantom described in the errata of \cite{kak2001principles}. 
    This figure is for illustrative purposes; the effects of the regularisations may vary with the measurement dataset.
    \textbf{From left-to-right, top-to-bottom:} each consecutive image illustrates the effect of adding an additional regularisation technique. All reconstructions used the DC correction strategy described in \S\ref{sec:discretising_fourier_filtration}.
    }
    \label{fig:shepplogan_ablation}
    \end{centering}
\end{figure}

In practice, reconstructions are performed on discrete measurement data to produce tomograms on finite rectilinear voxel lattices. A na\"ive application of the continuum theory to the discrete domain results in significant discretisation error that greatly hampers the quality of the reconstructed tomograms. This can be seen in Figure~\ref{fig:shepplogan_ablation} in the ``no regularisation'' subfigure. The discretisation errors from applying the continuum GBC formula na\"ively are so severe that the low-contrast features of the object are totally invisible. Adaptations are required before GBC can be practically applied to discrete datasets.

In this article, we propose four \emph{discretisation regularisations}, which are error-quelling modifications to the na\"ive implementation. We call them \emph{regularisations} because they preserve the continuum limit of the inversion formula (i.e. the limit in which there is an infinitely large reconstruction domain, with infinitely small voxels, and infinitely many projections).
We describe these regularisations as they are applied to the cylindrical source locus, but the concepts generalise readily to other multidimensional loci.
The cylinder serves as our prototype, and example reconstructions from the resulting regularised formula will be presented in \S\ref{sec:demonstrations}.

\subsection{Necessary number of source points for tomographic reconstruction on a voxel lattice}
\label{sec:sufficiency}
Before we discuss the discretisation regularisations, we will provide a Crowther-type \cite{crowther1970reconstruction} rule of thumb for estimating the number of X-ray source points required to stably saturate the Nyquist limit of the voxelised volume. This approach is similar to that put forth in \cite{kingston2018space}.

When the reconstruction domain is discretised into a 3D lattice of finitely many cube-shaped voxels, then there is theoretically a threshold number of projections beyond which there is sufficient data to exactly reconstruct the volume. 
(To see this, one may think of the tomographic reconstruction problem in its most basic form as a finite system of linear equations. With enough independent equations, the solution is uniquely determined.) 
However, the continuum problem may be analysed using a Crowther-type sampling criterion to produce an approximate number of required source points to saturate the Nyquist limit associated with the voxelised volume. This results in an estimate for the number of source points required to stably resolve the discretised volume up to its Nyquist limit. From this we produce an approximate `sufficiency' requirement. This is to be understood as an approximate guideline for stably reconstructing the object at a given resolution, rather than a strict data sufficiciency based on the discrete formulation of the problem.

We state here the approximate number of source points required to stably reconstruct from a cylindrical locus. We make these assumptions:
\begin{itemize}
    \item the detector is square,
    \item the detector has a sufficiently high pixel resolution, 
    \item the voxels are cubic, with sidelength $w$, and 
    \item the reconstruction support is a cylinder of maximal radius: $R_V = R \sin \vartheta_h$.
\end{itemize}
Under these assumptions, the number of source points should be equal to the height of the source cylinder, measured in voxel-lengths, multiplied by $\Lambda_z$, where 
\begin{equation*}
    \Lambda_z \gtrsim \pi \sec \vartheta_h \, . \tag{see \eqref{eq:data_sufficiency_approx_02}}
\end{equation*}
More explicitly, with a reconstruction support of height $h$ and radius $R_V$, the number of source points required, $m$, is
\begin{equation}\label{eq:data_sufficiency_req}
    m \gtrsim \pi \sec(\vartheta_h) \left. \frac{1}{w} \middle( h + 2(R_V+R) \tan(\vartheta_v) \right) \, . 
\end{equation}
A similar formula for rectangular detectors is found in the appendix \ref{sec:appendix_sufficiency}. Practical tomography rarely meets this sufficiency threshold: an example is provided in \S\ref{sec:demonstrations}.

\subsection{The `Unregularised' Discrete Theory}
We describe here what we consider to be the ``unregularised''\footnote{
    There is no single `correct' translation of the continuum theory on $\mathbb R^3$ to the domain of the finite voxel lattice; the problem is fundamentally ambiguous and is only constrained by the requirement that the discrete formula approach the continuum formula in the appropriate limit. We choose the most `natural' discrete formula as the starting point (and call this the ``unregularised'' formula), and layer discretisation regularisations on top that preserve the correct continuum limit while correcting for discretisation error away from that limit.} 
discrete implementation of the continuous inversion formula \eqref{eq:cone_beam_inversion_formula}. The vector $\mathbf m$ of measurements is now understood to be of finite length, encoding a finite number of integrated line attenuations through the volume, corresponding to the finite number of detector pixels and source positions. The weighted backprojection now sums over a finite set of source points rather than integrating over the idealised locus. The volume space, $\mathbb R^3$, is replaced with a finite 3D rectilinear lattice of voxels. The continuous Fourier transform, $F_{\mathcal A}$, is replaced with the 3-dimensional Discrete Fourier Transform of the voxel lattice, $\mathrm{DFT}$.\footnote{
    To resolve a subtle technical ambiguity related to the periodicity of the Discrete Fourier Transform: we use the conventional domain of the frequencies $\xi$ centered on $0$.
} 
The tomogram, $\mathbf x$, is now a tomogram on the voxel lattice. The unregularised discrete inversion formula is
\begin{subequations}\label{eq:cone_beam_inversion_formula_discrete_unregularised}
\begin{align}
    \mathbf x 
    \!&=\! \underbrace{\mathrm{DFT}^{-1} \; T \; \mathrm{DFT}}_{\text{cyclic deconvolution}} \underbrace{\left( \sum_x W_x B_x \right)}_{\text{weighted backprojection}} \mathbf m 
\end{align}
where
\begin{align}
	T &= \mathrm{diag}_{\xi}\left( \big. 
        {|\xi|} / \mathcal I[f](\hat{\xi})
	\!\right), &
W_x &= \mathrm{diag}_r\left( \big. 
	W\left(x, r\right) 
	\right) 
    . \label{eq:cone_beam_inversion_formula_discrete_unregularised_T}
\end{align}
\end{subequations}
The Funk transform $\mathcal I[f]$ and the weighting function $W(x,r)$ assume their continuum values given in Equations~\eqref{eq:cyl_eqs_I} and \eqref{eq:cyl_eqs_weights} respectively. 
The operator $\mathrm{diag}_r(\dots)$ multplies a spatially-varying function by the expression $(\dots)$ which is a function of the position $r$ in space. The centre positions of voxels are used for $r$. The operator $W_x B_x$ has a simple geometric interpretation: the value of the $i^{\text{th}}$ voxel, $(W_x B_x \mathbf m)_i$, is equal to the attenuation measured along the ray beginning from $x$ and intersecting the centre point $r_i$ of the voxel, multiplied by $W(x, r_i)$. For our purposes, the attenuation measured along that line is determined by linear interpolation between measurements on the pixelated detector. Also, $T$ is taken to multiply the $\xi = 0$ frequency by $0$.

\subsection{Backprojection antialiasing}
\label{sec:backproj_softening}
The expression for the backprojection weighting, $W(x,r)$, was given in \eqref{eq:cyl_eqs_weights} in terms of the angles $\theta_{\text{el}}$ and $\theta_h$. The purpose of this weighting is to ensure that all points $r$ in space receive backprojections from the same range of angles $\Omega$, and at a uniform density over $\Omega$. The masking out of backprojections with $|\theta_{\text{el}}| > \vartheta_v$ is the same as the Colsher window in \cite{colsher1980fully}.
However, because the detector and volume are discrete, this masking produces aliased edges in the backprojection, due to the sudden transition to $W(x,r) = 0$ on the edge of $\Omega$. 
Aliased edges can be catastrophic to the reconstruction quality because edges are highly sensitive to the volume deconvolution in \eqref{eq:cone_beam_inversion_formula_discrete_unregularised_T}. The deconvolution magnifies spatial frequency components in proportion with their frequency, $|\xi|$. Discontinuities in the backprojected volume contain infinitely high-frequency components, and so any aliasing of these edges produces severe artefacts in the reconstruction, on account of discretising the problem from the continuum. For an illustration of the effects, see the ``no regularisation'' reconstruction in Fig.~\ref{fig:shepplogan_ablation}; the artefacts are so severe that none of the object features are visible. 
We thus observe that a modification is needed to the reconstruction algorithm before it can be applied successfully to the discretised domain. 

\begin{figure}
\begin{center}
    \subfloat[]{
    {%
    \setlength{\fboxsep}{0pt}%
    \setlength{\fboxrule}{1pt}%
    \fbox{\includegraphics[width=0.48\linewidth]{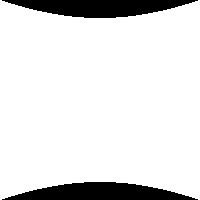}}%
    }%
    \label{fig:colsher_hardmask}
    }%
    \hfil
    \subfloat[]{
    {%
    \setlength{\fboxsep}{0pt}%
    \setlength{\fboxrule}{1pt}%
    \fbox{\includegraphics[width=0.48\linewidth]{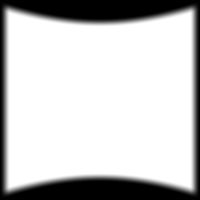}}%
    }%
    }
\end{center}
\caption{The standard Colsher window compared with the anti-aliased mask $s(\vartheta_v, \theta_{\text{soft}}; |\theta_{\text{el}}|) s(\vartheta_h, \theta_{\text{soft}}; |\theta_h|)$. Note that while anti-aliasing could be applied to the measurements as implied by these pictures, we actually apply the weightings to the voxels during backprojection. \textbf{(a):} the Colsher window \cite{colsher1980fully}. \textbf{(b):} The anti-aliasing mask.}
\label{fig:colsher_masks}
\end{figure}

We add a gradual windowing toward the edges of $\Omega$. This amounts to an additional `softening factor' as a funtion of $\theta_{\text{el}}$. This additional `softening factor' is illustrated in Figure~\ref{fig:colsher_masks} as a mask on the detector which is a `softer' variant of the hard-edge Colsher window. This softening alleviates the aliasing that occurs when the masked edge of detector measurements are backprojected into a discrete voxel lattice, thereby reducing associated artefacts. 
Adding this softening is not an approximation: we account for it exactly in the reconstruction formula by adding the same factor to $f$. This is necessary to avoid the introduction of new artefacts. With this addition, it is no longer true that $f = 1$ across all $\Omega$, and so the Funk transform $\mathcal I[f]$ must be recalculated. Care is needed in choosing the softening function so that it remains possible to compute $\mathcal I[f]$ in closed form. The details of the softening function, $s(\vartheta_v, \theta_{\text{soft}} ; \theta_{\text{el}})$, are contained in appendix \ref{app:exact_soften}. The angle $\theta_{\text{soft}}$ is the angular span over which the softening takes place. A typical choice for us is $0.05 \, \mathrm{rad}$.

Less importantly, the horizontal edges of the backprojection should also be soft-masked. Ideally, the object would fit entirely within the width of the detector and so the measurements at the horizontal extremities of the detector should be zero anyway, but in practice this is not the case because of measurement noise. Therefore, we apply a similar softening factor as a function of the horizontal angle $\theta_h$. The radius of the reconstruction support reduces to $R_V = R \sin(\vartheta_h - \theta_{\text{soft}})$.

The effect of adding this regularisation (`backprojection anti-aliasing') is seen in Fig.~\ref{fig:shepplogan_ablation}. Clearly, there is an improvement, but further regularisations are needed.

\subsection{Accumulated weight normalisation}
\label{sec:weights_normalisation}
There are reconstruction artefacts introduced by the discreteness of the source locus. These artefacts can appear in the volume as `winding' or `crosshatched' or `patchy' regions of low/high attenuation. These are visible in Fig.~\ref{fig:shepplogan_ablation}, before accumulated weight normalisation is applied. These artefacts are similar in nature to those encountered in other exact reconstruction methods when an incorrect choice of `\emph{windowing function}' is applied to the detector. 
The source of these artefacts is the fact that different points in the volume receive different (weighted) total numbers of backprojections. This is an inevitable consequence of approximating a continuous source locus with a set of discrete source points.

Other reconstruction methods are designed with specific, structured loci in mind, and implicitly resolve this issue due to the use of a windowing function on the detector that interacts with the regular structure of the locus to ensure that all points in the volume receive an exactly equal number of backprojections.
However, our aim is to develop a general method of inversion for multidimensional loci that imposes minimal requirements on the locus. To that end, we describe here a procedure, which we call \textit{accumulated weight normalisation}, that applies to the weighted backprojection to ensure that all points in the volume receive an equal number of backprojections. This reduces associated artefacts. This procedure is more computer-memory intensive than simple windowing, but it is also completely generic (it does not depend on the choice of locus).

The procedure is outlined as follows: as the weighted backprojection $(\sum_x W_x B_x) \mathbf m$ is calculated, we record in parallel another volume into which the softened weights are accumulated: $S_A = \sum_x W_x$. After the backprojection is performed, it is normalised by comparing $S_A$ with the \emph{expected softweights}: $S_E = \int_{\mathbb R^3} \mathrm dx \chi(x) W_x$.
$S_E$ is computed by numerical integration over the idealised source point distribution. Symmetries in source loci such as the cylinder or sphere can be exploited to speed up the computation of $S_E$.
\textit{Accumulated weight normalisation} is then applied to the weighted backprojection as follows:
define $Z(\epsilon; a) = a + \epsilon e^{-a/\epsilon}$.
Then
\begin{enumerate}
    \item Set $S_A := Z(10^{-6}; S_A)$. (This replaces $0$s with small positive numbers.)
    \item Multiply the backprojection by $S_E/S_A$.
\end{enumerate}

The effect of adding this regularisation (`accumulated weight normalisation') is seen in Fig.~\ref{fig:shepplogan_ablation}. Low-frequency and high-frequency artefacts remain, and these warrant further regularisations.

\subsection{Discrete Fourier Correction}
\label{sec:discretising_fourier_filtration}
The deconvolution in \eqref{eq:cone_beam_inversion_formula} was derived from the continuous theory, in continuous space $\mathbb R^3$. In practice, we work with discretised volumes that are split into a finite number of voxels. 
Attempting to apply a deconvolution to a discrete volume by applying the same multiplier $\mathrm{diag}(...)$, as in \eqref{eq:cone_beam_inversion_formula_discrete_unregularised_T}, is mathematically incorrect (a similar point has been made in \cite{zeng2014revisit}) and results in high-frequency artefacts that are visible in Fig.~\ref{fig:shepplogan_ablation} before the discrete Fourier correction is applied. 
We dissect this matter in more detail in the appendix \ref{app:discrete_fourier_filtration}. Stemming from that analysis, our recommendation (which is only an approximation) is to add the following additional factor to the transfer function $T$ in \eqref{eq:cone_beam_inversion_formula_discrete_unregularised}:
\begin{equation}
\mathrm{sinc}\left(\pi l \xi_x\right)
\mathrm{sinc}\left(\pi l \xi_y\right)
\mathrm{sinc}\left(\pi l \xi_z\right) \, ,
\end{equation}
where $l$ is the side length of the cubic voxels. The $\mathrm{sinc}$ function is defined in \eqref{eq:sinc_function}.

Through various simulated test reconstructions, we find that this approximation is highly effective in suppressing high-frequency artefacts that otherwise surface from the na\"ive application of the deconvolution filter in \eqref{eq:cone_beam_inversion_formula_discrete_unregularised}. It also does not appear to introduce any new artefacts.
See Fig.~\ref{fig:shepplogan_ablation}.

It is also necessary to make a correction to the zero-frequency component of the reconstruction, due to the inexactness of our discretisation of the Fourier filtration (cf. \cite{crawford1991}). This may be accomplished by observing the average value of the reconstruction within voxels which are known to be $0$ from a priori knowledge of the object. For our simulated examples, we accomplished this by sampling the top and bottom z slices of the backprojected volume, recording which voxels were $0$, and using the mean value of these voxels as the $0$ point after the convolution. For the experimental example, we set the zero-frequency component of the reconstruction to $0$ and performed a manual correction afterwards. 

In equation form, this and the previous regularisations enhance the discrete inversion formula \eqref{eq:cone_beam_inversion_formula_discrete_unregularised} to the following:
\begin{subequations}\label{eq:recon_discrete_reg_03}
\begin{align}
    &\mathbf x 
    \!=\! \underbrace{\mathrm{DFT}^{-1} \; T \; \mathrm{DFT}}_{\text{cyclic deconvolution}} \underbrace{
    \left( \frac{\int_{\mathbb R^3} \chi(x) W_x }{Z\left( \epsilon ; \sum_x W_x \right)} \right)
    }_{\text{accum. weight. norm.}}  \underbrace{\left( \sum_x W_x B_x \right)}_{\text{weighted backprojection}} \mathbf m \nonumber \\ 
    & \qquad \qquad + \text{DC\_correction}
\intertext{where} 
&W_x = \mathrm{diag}_r\!\left( \big. 
        W(x,r) 
        \underbrace{
            s\!\left(\vartheta_v, \theta_{\text{soft}} ; \theta_{\text{el}} \right)
            s\!\left(\vartheta_h, \theta_{\text{soft}} ; \theta_h \right) 
        }_{\text{backprojection anti-aliasing}}
        \right) \\
&T = \mathrm{diag}_{\xi}\!\Bigg(\! \big. 
        \underbrace{
            \mathrm{sinc}\!\left(\!\pi l \xi_x\!\right)
            \mathrm{sinc}\!\left(\!\pi l \xi_y\!\right)
            \mathrm{sinc}\!\left(\!\pi l \xi_z\!\right) 
        }_{\text{discrete Fourier correction}}
        \underbrace{
            \frac{|\xi|}{G(\vartheta_v, \theta_{\text{soft}}; \xi)}
        }_{\text{backproj. A.A.}}
    \!\Bigg) \\
&Z(\epsilon; a) = a + \epsilon e^{-a/\epsilon} \, ,
\end{align}
\end{subequations} 
where $W(x,r)$ is given in \eqref{eq:cyl_eqs_weights}, $s(a,b; c)$ is given in \eqref{eq:exact_softening_function}, $G(\vartheta_v, \theta_{\text{soft}}; \xi)$ is given in \eqref{eq:closed_form_integral}, and $l$ is the physical side length of the cubic voxels. The ``DC\_correction'' was described in the previous paragraph.

There remain low-frequency artefacts which are corrected by a final regularisation.

\subsection{Low-pad correction}
\label{sec:lowpad_correction}
Theoretically, an infinitely large volume is required in order to perform the exact reconstruction using \eqref{eq:cone_beam_inversion_formula}. That's because the deconvolution kernel is not compactly supported, i.e. it has `infinite range' (every point in the tomogram depends on backprojection data from an infinite distance away). Therefore, the backprojection should be computed within an infinitely large volume, and only then should the deconvolution be applied.
This is obviously not possible to perform on a discrete voxel lattice with finite bounds. 
In practice, using a large-but-finite padding on the reconstruction support produces tomogram artefacts that are only of a low spatial-frequency (see Fig.~\ref{fig:shepplogan_ablation} before and after the low-pad correction). A similar concept appears in a different reconstruction method \cite{ye2005general}, wherein `PI-lines' must be padded during reconstruction.

We supply a fixed padding factor to every axis of the volume, and after the tomogram is produced within this padded region, we perform a \textit{low-pad correction} to remove the majority of the remaining error that is induced by a lack of padding. The low-pad correction term is the residual between a reconstruction with a large amount of padding and one with a small amount of padding. Because it is a low-spatial-frequency effect, the low-pad correction term can be probed at a lower resolution, making this regularisation computationally affordable.

Our \text{low-pad correction} technique is comprised of 
\begin{itemize}
\item a weighted backprojection into a secondary volume (the lower resolution reconstruction) with voxel dimensions $n$ times larger in each axis but with a large padding factor $f_2$. This backprojection is performed in parallel with the backprojection into the full-resolution volume that has a smaller padding factor $f_1$. 
\item performing the reconstruction on the secondary volume.
\item cropping the unfiltered secondary volume so that it has an identical padding proportion to the full-resolution volume.
\item performing the reconstruction on the cropped secondary volume.
\item comparing the difference between the cropped and uncropped reconstructions on the secondary volume.
\item adding this difference to the full-resolution volume, via trilinear upscaling. 
\end{itemize}
When we say ``performing the reconstruction'' above, we are including all other regularisations previously described.
Suggested values of $f_1$, $f_2$ and $n$ are $f_1 = 1.2$, $f_2 = 6$, and $n = 9$. A padding factor of $f_1 = 1.2$ indicates that the reconstruction domain should be padded with zeros by $10\%$ of its length on each side of each axis, and a downscaling factor $n = 9$ indicates that each voxel in the secondary volume should correspond to a block of $9 \times 9 \times 9$ voxels of the full-resolution volume. 

If applicable, the padding factor for the $z$ axis (which aligns with the cylinder axis) is clipped to a maximum value so that the backprojected volume is not everywhere zero at the $z$ extremities. This is because additional padding on the $z$ axis is unnecessary. 

\section{Demonstrations on Cylindrical Loci}\label{sec:demonstrations}

In the previous sections, we summarised the general theory of global backprojection-convolution (GBC) (\S\ref{sec:inversion_theory}) 
and introduced practical discretisation regularisations (\S\ref{sec:practical_discrete_methods}) to implement GBC on realistic (i.e. discretised) measurement data. 
In this section, we present a simulation study in \S\ref{sec:sims_01} and a brief experimental validation in \S\ref{sec:experimental_recon}.

Relevant hardware used for GBC reconstruction: 
a consumer personal computer with the following relevant specifications: GPU: NVIDIA GeForce RTX 3090 24GiB; CPU: AMD Ryzen 9 3950X (32-core) @ 3.5GHz; RAM: 4 x 16GB DDR4 @ 3200 MT/s.

\subsection{Simulation study: 3D FORBILD head phantom}
\label{sec:sims_01}

\subsubsection{The ground truth data}
\label{sec:sims_01_gt_desc}
The ground truth volume used for generating the projection dataset was the 3D FORBILD head phantom \cite{forbild}. This phantom is challenging to reconstruct due to the high-contrast, high-frequency features which produce artefacts that tend to swamp out the low-contrast features. 

\subsubsection{The simulated source locus}
\label{sec:ldt}

For this simulation, we used a \emph{low-discrepancy sequence} (LDS) of source points on the cylinder \cite{priv_conv_LDS}. LDSs are in some sense optimal as infinite sequences of points that uniformly spread across some continuous space. (See \cite{mohan2015timbir} for an application of a 1D LDS to dynamic tomography.) This makes them suitable for testing the correctness of the proposed inversion algorithm. 
LDSs also have the advantage that any contiguous subsequence of the LDS is also an LDS, and therefore also approximately uniformly samples the space. This is convenient for simulation purposes, but we would not suggest the direct use of an LDS for real imaging with a single X-ray source, due to the inefficiencies in moving the source between consecutive points of the sequence.

\begin{figure}[!ht]
    \begin{center}
    \includegraphics[width=\columnwidth]{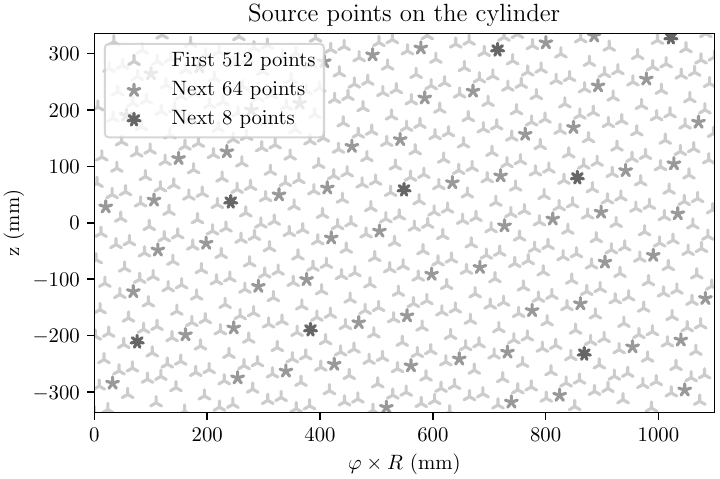}
    \vspace*{-0.7cm}
    \end{center}
    \caption{An illustration of the low-discrepancy sequence (LDS).
    }
    \label{fig:ldt}
\end{figure}

The cylindrical source locus has been approximated with an LDS as follows. The cylinder has a periodic azimuthal coordinate $\varphi$ and a $z$ coordinate (aligned with the cylinder axis) with finite bounds. We use the algorithm described in \cite{roberts2018lds} that produces an LDS on $n$-dimensional unit cubes. In order to adapt the 2D unit cube (unit square) to the source cylinder of height $h = (z_{\text{max}} - z_{\text{min}})$ and radius $R$, we parameterise the cylinder by coordinates 
\begin{align*}
    x_1 &= \frac{z - z_{\text{min}}}{\max\left\lbrace h \, , 2 \pi R \right\rbrace} , \quad
    x_2 = \frac{\varphi R}{\max\left\lbrace h \, , 2 \pi R \right\rbrace} \, .
\end{align*}
We then define the sequence
\begin{align*}
(x_1, x_2)_i = (i \phi_2^{-1} - \lfloor i \phi_2^{-1} \rfloor, i \phi_2^{-2} - \lfloor i \phi_2^{-2} \rfloor) \quad \text{for} \quad i \in \mathbb N \, ,
\end{align*}
where $\phi_2 \approx 1.3247\dots$ is the unique positive real solution to $\phi_2^{3} = \phi_2 + 1$ (see \cite{marohnic2012plastic} for `harmonious numbers') and $\lfloor \cdot \rfloor$ rounds down to the nearest integer.
We then filter out any points of the sequence that fall outside of the cylinder range; that is, we reject elements of the sequence for which either
\begin{align*}
    x_1 &> \frac{h}{\max\left\lbrace h \, , 2 \pi R \right\rbrace} , \quad
    \text{or} \quad x_2 > \frac{2 \pi R}{\max\left\lbrace h \, , 2 \pi R \right\rbrace} \, .
\end{align*}
Figure~\ref{fig:ldt} depicts the resulting locus.

\subsubsection{Production of the simulated projection data}
\label{sec:sims_01_prod_proj_data}
Projection data was generated from the ground truth, without noise, by computing exact line integrals of the attenuation coefficient through the primitive shapes that comprise the phantom (e.g. cylinders, ellipsoids). Each line integral was computed from the X-ray source to the centre of the corresponding detector pixel.
\begin{samepage}
\begin{center}
FORBILD head phantom data generation parameters\newline
\begin{tabularx}{0.98\linewidth}{|X|c|}
\hline
source cylinder radius ($R$) & $\approx175\;\mathrm{mm}$ \\
\hline
source cylinder height & $\approx 673\;\mathrm{mm}$ \\
\hline
source-to-detector dist. ($L$) & $\approx 425\;\mathrm{mm}$ \\
\hline
detector size (pixels)   & $2001 \times 2001$ pixels \\
\hline
detector size (physical) & $\approx 850\;\mathrm{mm} \times 850\;\mathrm{mm}$ \\
\hline
$\vartheta_h$ / $\vartheta_v$ & $\approx \! 45.00 \!$ deg / $\approx \! 35.26 \!$ deg \\ 
\hline
number of source points & up to $64000$ \\
\hline
\end{tabularx}
\end{center} 
\end{samepage}
The source cylinder height ($\approx 673\;\mathrm{mm}$) is sufficiently large to capture all relevant projections of the object. The X-ray cone angles $\vartheta_v, \vartheta_h$ have been chosen large to demonstrate that the reconstruction method works with arbitrary cone angles.

\subsubsection{Reconstruction}\label{sec:sim_recon_subsec}
See the below table for information about the reconstruction size and parameters.
\begin{samepage}
\begin{center}
FORBILD head phantom reconstruction parameters\\
\begin{tabularx}{0.98\linewidth}{|X|c|}
\hline
recon. voxel width & $0.3\;\mathrm{mm}$ \\
\hline
recon. size (voxels) & $837 \times 837 \times 837$ \\
\hline
recon. support cylinder radius ($r$) & $\approx 124\;\mathrm{mm}$ \\
\hline
recon. support cylinder height ($h$) & $\approx 250\;\mathrm{mm}$ \\
\hline
\hline
recon. formula used & \eqref{eq:recon_discrete_reg_03} plus \S\ref{sec:lowpad_correction} \\
\hline
formula parameters $f_1, f_2, n, \epsilon, \dots$ & $1.2, 6, 9, 10^{-6}$\\
\hline
$\dots \theta_{\text{soft}}$ (horz.), $\theta_{\text{soft}}$ (vert.) & $0.025, 0.100\;\mathrm{rads}$ \\
\hline
estimated no. of source points required for data sufficiency & $\approx 9963$, using \eqref{eq:data_sufficiency_req}\\
\hline
\end{tabularx}
\end{center} 
\end{samepage}

We reconstructed several tomograms from the data by drawing from the first $m$ source points of the low-discrepancy sequence locus, where $m$ was varied. This gives us a sense of how the algorithm performs on variously sparse/fine distributions of source points on the cylinder, approaching the continuum limit $m \rightarrow \infty$ in which the inversion theory is founded.

\subsubsection{Results}
\begin{figure}[ht!]
    \begin{center}
    \includegraphics[width=0.99\linewidth]{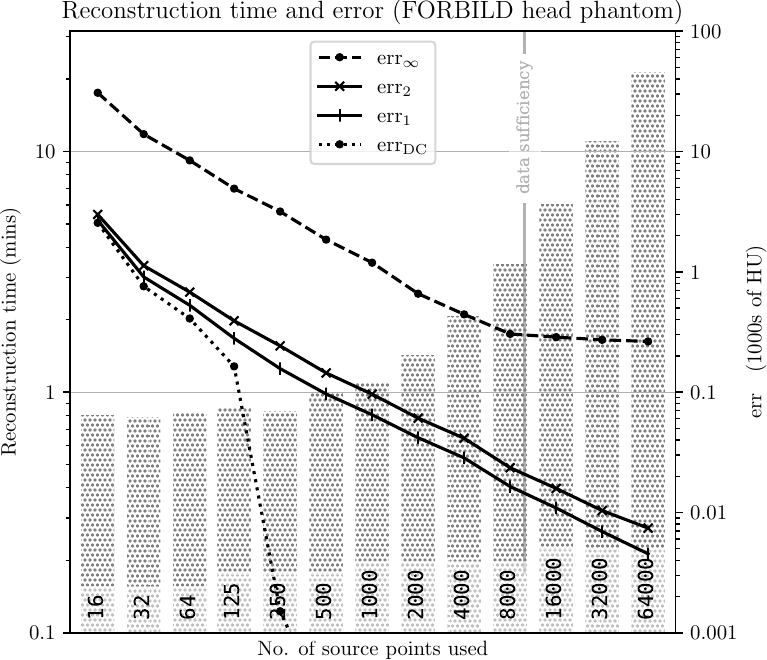}
    { \small
    \vspace*{-0.5cm}
    \begin{align*}
        \text{err}_\infty &= {\max\left\lbrace |\text{reconst.} - \text{groundtruth}| \right\rbrace}\, .\\
        \text{err}_2 &= \sqrt{{
            \mathrm{mean} \Big(|\text{reconst.} - \text{groundtruth}|^2 \Big)
        }}\, .\\
        \text{err}_1 &= {{
            \mathrm{mean} \Big(|\text{reconst.} - \text{groundtruth}| \Big)
        }}\, .\\
        \text{err}_{\text{DC}}
        &= \left| \mathrm{mean} \Big( \text{reconst.} - \text{groundtruth} \Big) \right| \, .
    \end{align*}
    \vspace*{-0.5cm}
    }
    \caption{
        A graph of reconstruction time and error for the simulated dataset referred to in \S\ref{sec:sims_01}.
        Indicated in the graph is a rough estimate of the number of projections required for data sufficiency, computed from \eqref{eq:data_sufficiency_approx_02}.
        Consult \S\ref{sec:sims_01_gt_desc} for details about the phantom.
        The $\mathrm{err}_1$ of the highest-quality reconstruction is seen here to be $\approx 4.5 \; \mathrm{HU}$. The $\mathrm{err}_2$ (RMSE) is $\approx 7.5 \; \mathrm{HU}$. Vacuum is defined as $-1000 \; \mathrm{HU}$ and water is defined as $0 \; \mathrm{HU}$.
    }
    \label{fig:err_moms_smallvolume_graphs}
    \end{center}
\end{figure}

\begin{figure*}
    \begin{centering}
    \vspace{2mm}
    \includegraphics[width=1.00\linewidth]{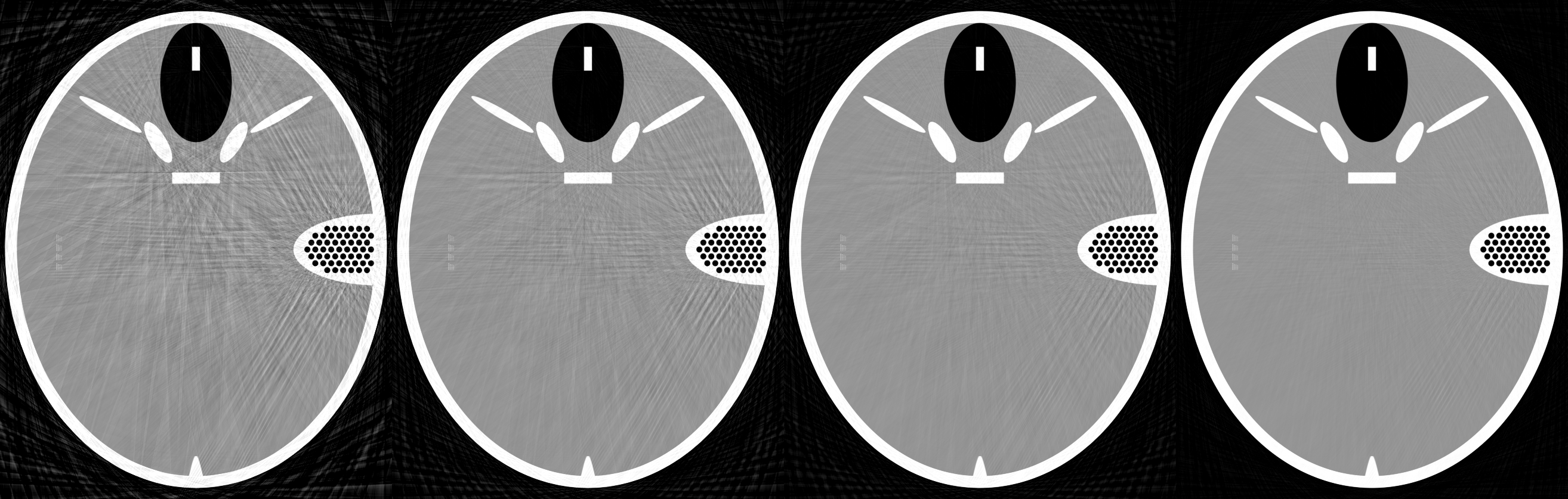}%
    \caption{Slices of reconstructions of the FORBILD head phantom. Plotting range: $[-1000,800] \; \mathrm{HU}$. The white regions are bone with attenuation coefficient $800 \; \mathrm{HU}$. The black regions are vacuum with attenuation coefficient $-1000 \; \mathrm{HU}$. Therefore, $\approx 100\%$ of the full range of attenuation coefficient is visible. \textbf{Left to right:} 1000 source points; 2000 source points; 4000 source points; 8000 source points. See \S\ref{sec:sims_01}.
    }
    \label{fig:recon_slices_a}
    \end{centering}
\end{figure*}

\begin{figure*}
    \begin{centering}
    \vspace{2mm}
    \includegraphics[width=1.00\linewidth]{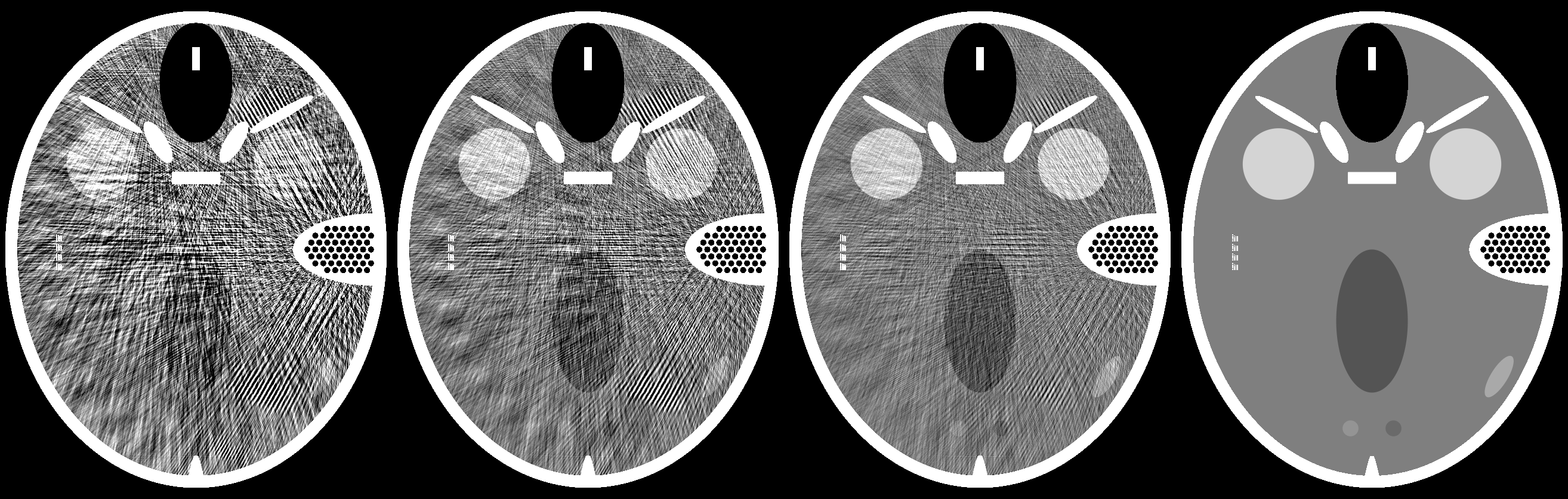}%
    \caption{Slices of reconstructions of the FORBILD head phantom. Plotting range: $[35,65] \; \mathrm{HU}$. The white regions are bone with attenuation coefficient $800 \; \mathrm{HU}$. The black regions are vacuum with attenuation coefficient $-1000 \; \mathrm{HU}$. Therefore, $\approx 1.7\%$ of the full range of attenuation coefficient is visible. \textbf{Left to right:} 16000 source points; 32000 source points; 64000 source points; ground truth. See \S\ref{sec:sims_01}.
    }
    \label{fig:recon_slices_b}
    \end{centering}
\end{figure*}

To assess the quality of the reconstructions, we require a ground truth. The ground truth was computed by sampling the FORBILD phantom at a resolution of $0.03\;\mathrm{mm}$, and then averaging the values within $10\times10\times10$ blocks, producing a voxelised ground truth with a resolution of $0.3\;\mathrm{mm}$, which matches the reconstructions. The reconstructions were then compared with this ground truth on a voxel-wise basis.
Figure~\ref{fig:err_moms_smallvolume_graphs} contains error metrics and reconstruction times.
Figures~\ref{fig:recon_slices_a} and \ref{fig:recon_slices_b} depict a cross-section of various reconstructions.
Note that ``$\mathrm{HU}$'' refers to `Hounsfield Units'. Vacuum is defined as $-1000\;\mathrm{HU}$ and water as $0\;\mathrm{HU}$.
 
\subsubsection{Discussion}
The $\mathrm{err}_1$ (i.e. average absolute voxel deviation) of the reconstructions diminished with the number of source points in an approximate power-law relationship. The $\mathrm{err}_{\infty}$ (i.e. the maximum deviation between the reconstruction and the ground truth) plateaued from around the number of source points predicted for sufficiency. 

The final reconstruction (using $64,000$ source positions) visibly shows the two small dots at the bottom, with attenuation coefficients of $52.5\;\mathrm{HU}$ and $47.5\;\mathrm{HU}$ against a background of $50.0\;\mathrm{HU}$ in the homogenous brain matter. These are the lowest-contrast features of this phantom.

The reconstructions are marred by strong streak artefacts emanating from the right `inner ear'. These artefacts are commonly encountered when reconstructing this phantom. In the final reconstruction ($64,000$ source points), the regions of greatest error were the high-contrast edges, particularly within the right `inner ear'.

Each reconstruction was performed with a single backprojection followed by a volume convolution. Computationally, this is comparable with $\tfrac{1}{2}$ to $1$ iteration of a conventional iterative method, which usually requires at least one projection and one backprojection of the dataset, per iteration. Iterative methods robust to noise usually require many iterations, whereas GBC directly minimises an objective function for the error metric in the continuum limit, and therefore always requires only a single iteration, even in the presence of noise. An experimental example including measurement noise is presented in the next section.


\subsection{Tomographic reconstruction from experimental data}
\label{sec:experimental_recon}
In this section, we demonstrate a reconstruction from experimentally-acquired data. The dataset is noisy and does not meet the theoretical requirement for sufficiency with the GBC method. This is because the range of $z$ coordinates over which the X-ray source is placed is insufficient. Also, the object extends axially beyond the source cylinder, which also violates an assumption of the GBC method.

\subsubsection{The ground truth data}
\label{sec:expt_01_gt_desc}
\begin{figure}[th!]
    \centering%
    \hfill%
    \subfloat[]{\raisebox{0.2\height}{\includegraphics[width=0.20\linewidth]{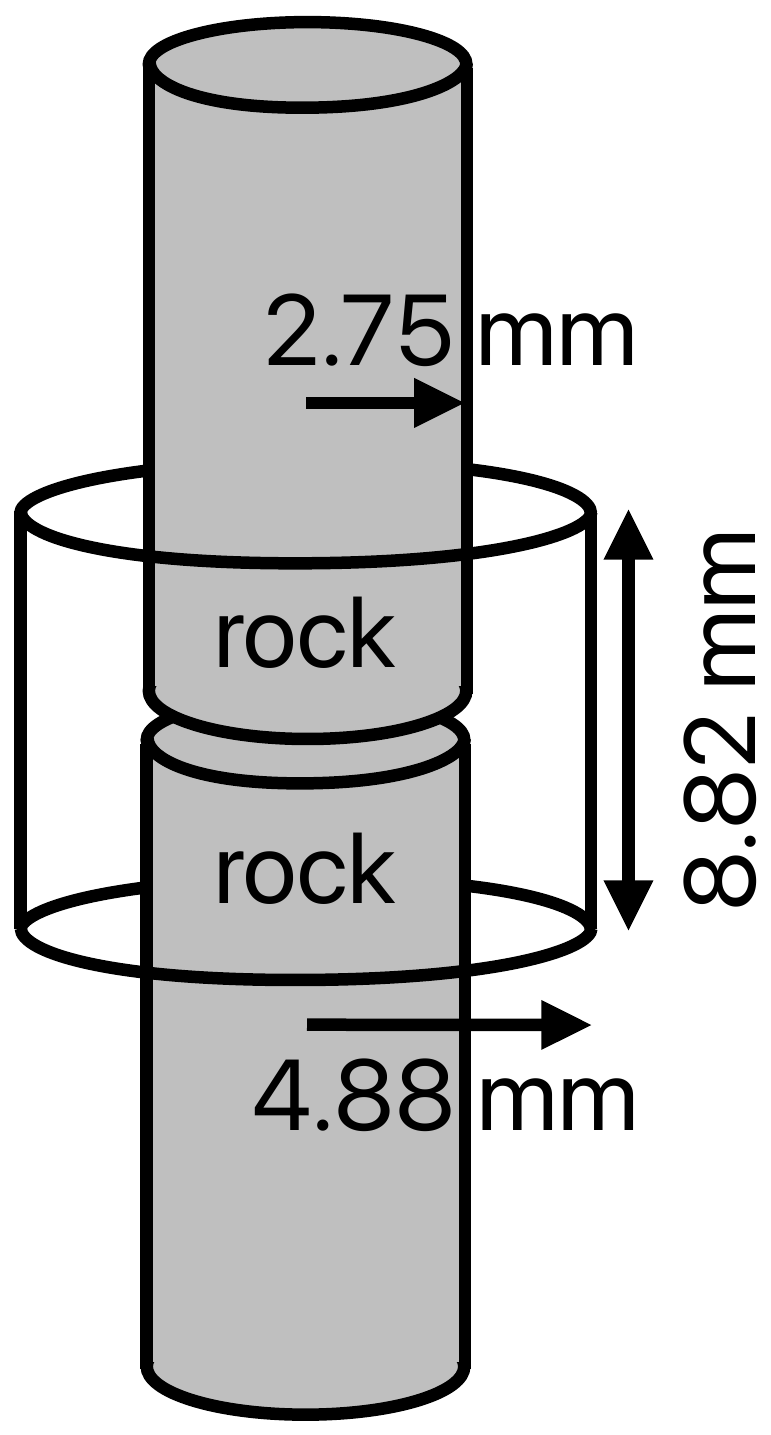}}}%
    \hfill%
    \subfloat[]{\includegraphics[width=0.52\linewidth]{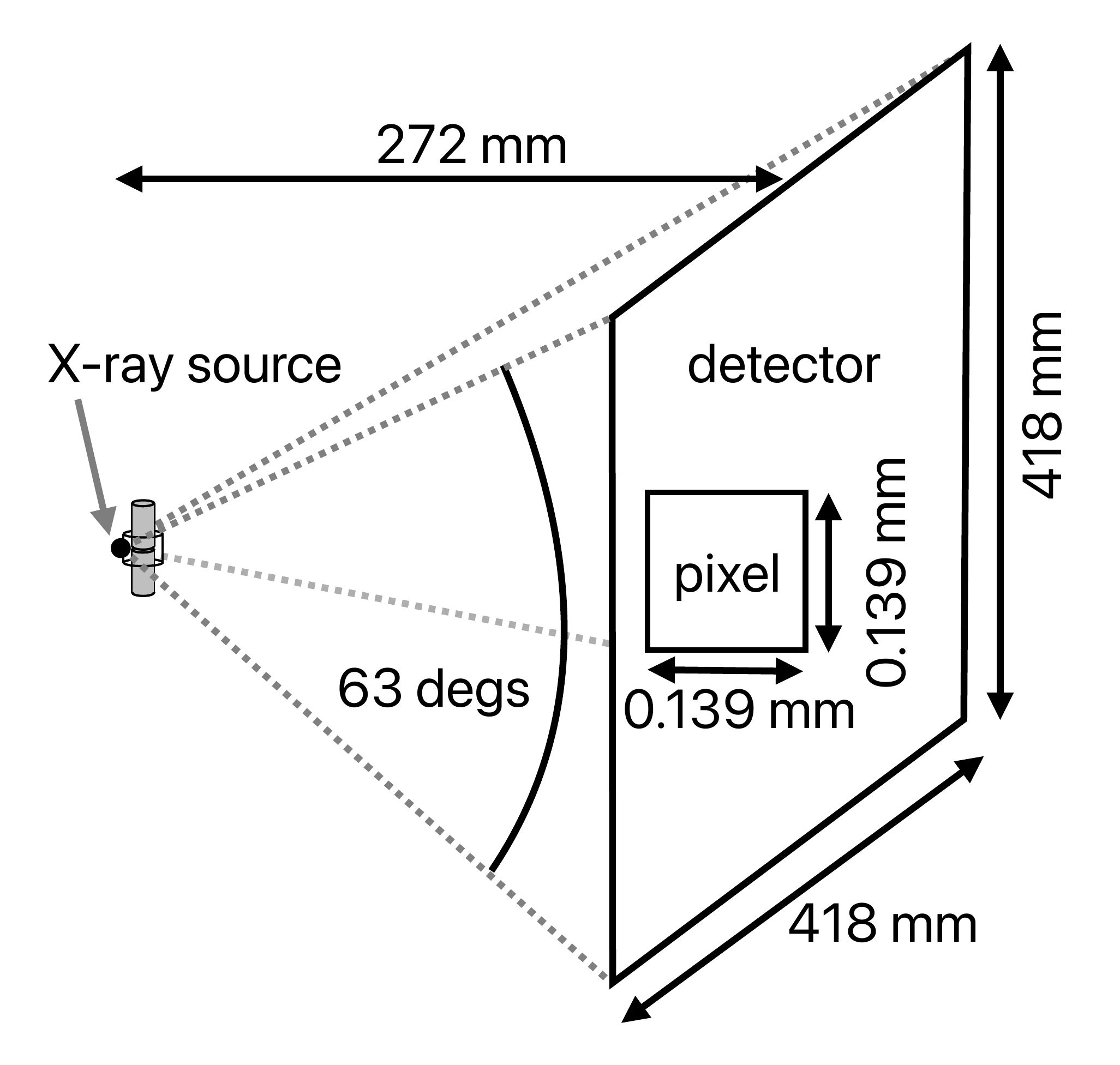}}%
    \hfill%
    \caption{Geometry of the experimental scans. The image magnification factor varies from $36$ (at the back of the sample) to $128$ (at the front of the sample).}
    \label{fig:exp_scan_geom}
\end{figure}

The volume scanned was a vertical stack of rocks taken from bored cylindrical `rock cores'. The scan was centered on the interface between two of these rocks. The geometry is depicted in Figure~\ref{fig:exp_scan_geom}. 

\subsubsection{The source locus}
\label{sec:expt_trajectory}
The locus used for this scan was an instance of the cylindrical space-filling trajectory (SFT) described in \cite{kingston2018space}. This is a low-pitch, sparsely-sampled helical trajectory that can be realised on any existing X-ray scanning apparatus supporting custom helical trajectories. It uniformly samples the source points on the cylinder. See Fig.\ref{fig:trajectories_sft} for an example with relatively few points.
For comparison with a well-known baseline, a double-helix (DH) scan was also performed.

\subsubsection{The projection data}
\label{sec:expt_projdata}
The essential geometric parameters of the scan data are given in the following table: 
\begin{samepage}
\begin{center}
Experimental data collection parameters
\begin{tabularx}{0.98\linewidth}{|X|c|}
\hline
object radius & $\approx 2.75$ mm \\
\hline
object height & $> 13$ mm \\
\hline
\hline
source cylinder radius ($R$) & $\approx 4.88$ mm \\
\hline
source cyl. height (DH, SFT)& $\approx 7.41\;\mathrm{mm}, \; \approx 8.82 \; \mathrm{mm}$ \\
\hline
$\Delta z$ between points (DH, SFT) & $\approx 1.60\;\mathrm{\mu m}, \; \approx 1.53 \; \mathrm{\mu m}$ \\
\hline
$\Delta \varphi$ between points (DH, SFT) & $\approx 0.100\;\mathrm{deg}, \; \approx 2.19 \; \mathrm{deg}$ \\
\hline
source-to-detector distance ($L$) & $\approx 272$ mm \\
\hline
detector size (pixels)   & $3040 \times 3040$ pixels \\
\hline
detector size (physical) & $\approx 418 \times 418$ mm \\
\hline
detector pixel size & $\approx 139 \times 139\;\mathrm{\mu m}$ \\
\hline
$\vartheta_h$ / $\vartheta_v$ & $\approx 37.5$ deg / $\approx 31.3$ deg \\
\hline
no. of source points (DH, SFT) & $2 \times 4622, \; 5754$ \\ 
\hline
exposure time per projection & $300\;\mathrm{ms}$ \\
\hline
\end{tabularx}
\end{center}
\end{samepage}

\subsubsection{Reconstruction}
\label{sec:expt_reconstruction}
Three reconstructions were performed: (1) a Katsevich-like filtered-backprojection method applied to the DH data; (2) a GBC reconstruction from the SFT data, with the parameters described above; (3) an iterative reconstruction from the SFT data, using the method described in \cite{myers2016rapidly}.

The parameters of the GBC reconstruction were as follows:
\begin{samepage}
\begin{center}
Experimental reconstruction parameters
\begin{tabularx}{0.98\linewidth}{|X|c|}
\hline
recon voxel size & $\approx 2.49 \times 2.49 \times 2.49 \, \mathrm{\mu m}$ \\
\hline
recon size (voxels) & $2385 \times 2385 \times 1377$ \\
\hline
recon size (physical) & $5.95 \times 5.95 \times 3.30$ mm \\
\hline
\hline
recon. formula used & \eqref{eq:recon_discrete_reg_03} plus \S\ref{sec:lowpad_correction} \\
\hline
alg. parameters $f_1, f_2, n, \epsilon, \dots$ & $1.2, 6, 9, 10^{-6}$\\
\hline
$\dots \theta_{\text{soft}}$ (horz.), $\theta_{\text{soft}}$ (vert.) & $0.050, 0.100 \; \mathrm{rads}$ \\
\hline
\end{tabularx}
\end{center}
\end{samepage}

\subsubsection{Results}
\begin{figure*}
\begin{centering}
    \begin{minipage}[c]{1.0\linewidth}
    \subfloat[double-helix; Katsevich; x-slice]{
    \hspace*{-0.2cm}
    \zoominset{0.322\linewidth}{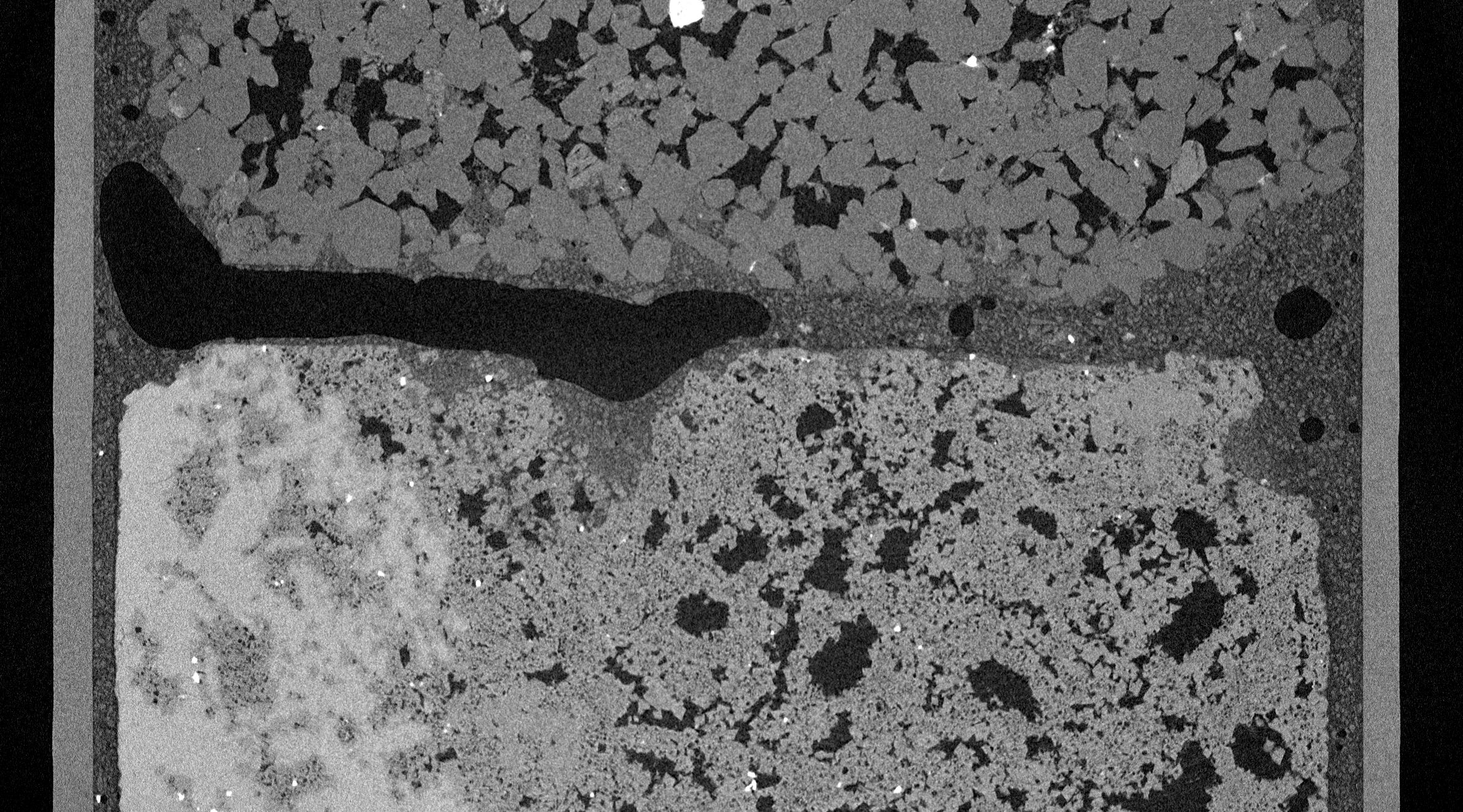}{4.5}{0.215,0.423}{0.315,0.523}{0.535,0.025}
    \vspace*{-0.5cm}
    }%
    \subfloat[SFT; GBC; x-slice]{
    \hspace*{-0.2cm}
    \zoominset{0.322\linewidth}{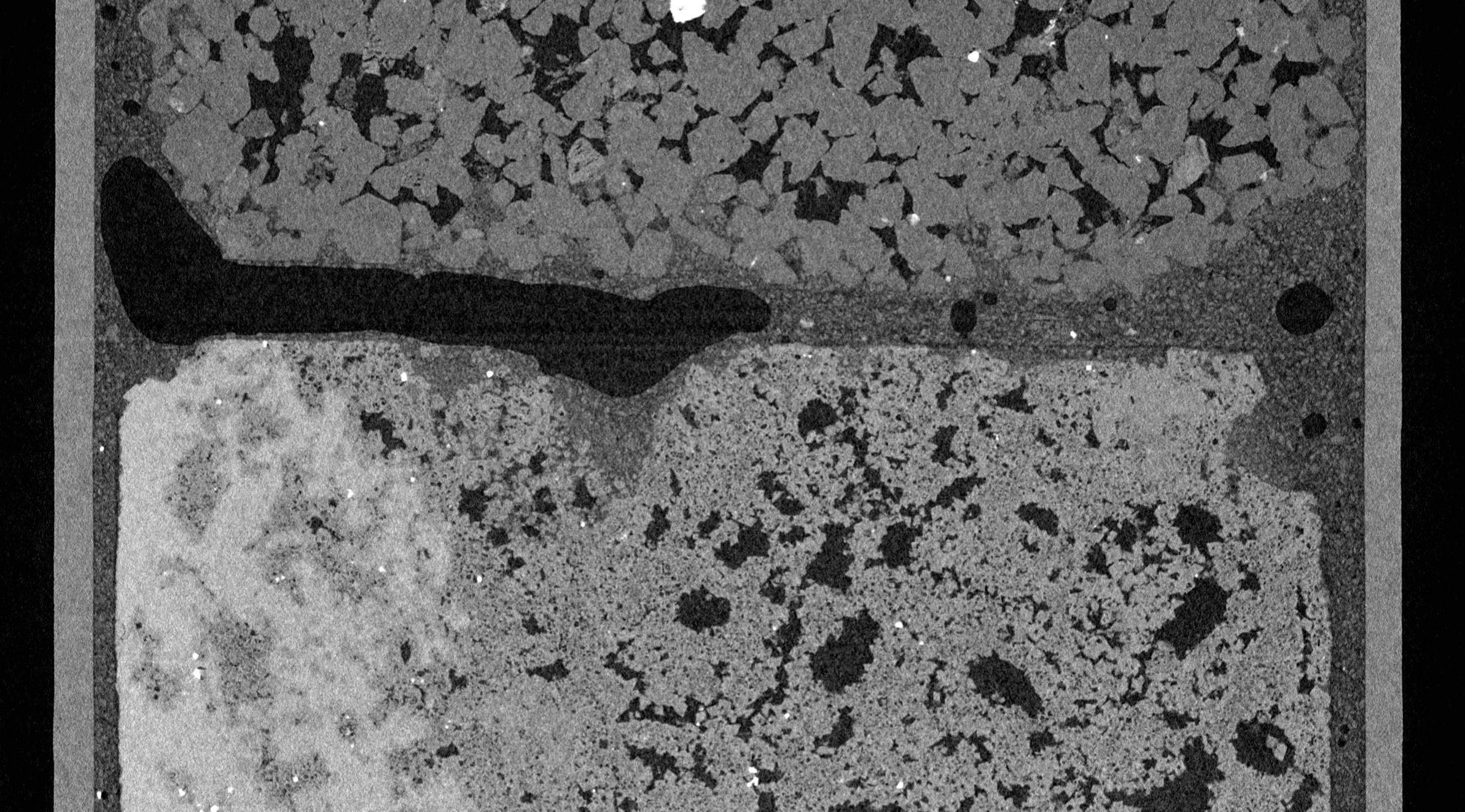}{4.5}{0.215,0.425}{0.315,0.525}{0.535,0.025}
    \vspace*{-0.5cm}
    }
    \subfloat[SFT; iterative; x-slice]{
    \hspace*{-0.2cm}
    \zoominset{0.322\linewidth}{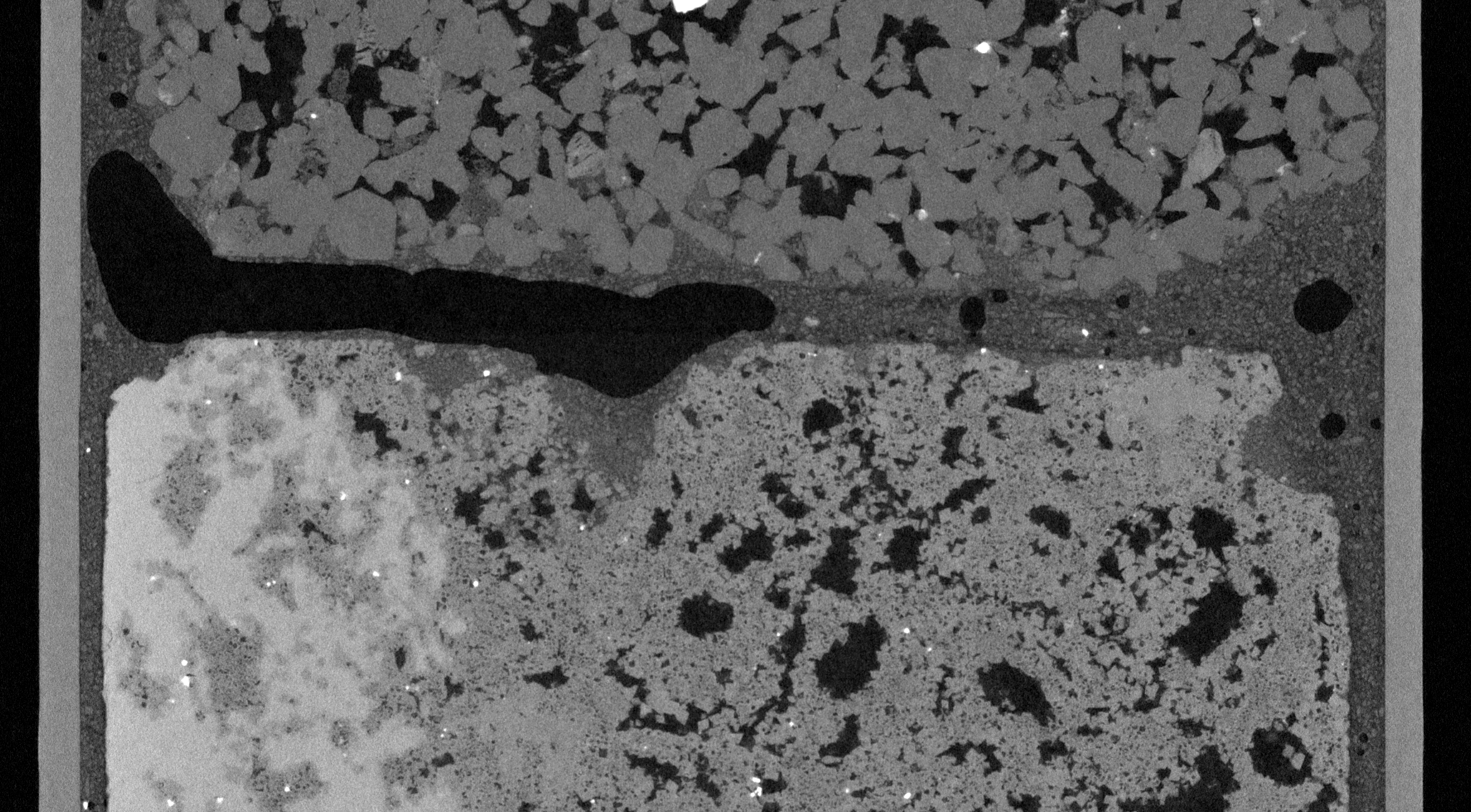}{4.5}{0.209,0.428}{0.309,0.528}{0.535,0.025}
    \vspace*{-0.5cm}
    }%
    \end{minipage}
    \newline
    \begin{minipage}[c]{1.0\linewidth}
    \subfloat[double-helix; Katsevich; z-slice]{
    \hspace*{-0.2cm}
    \zoominset{0.322\linewidth}{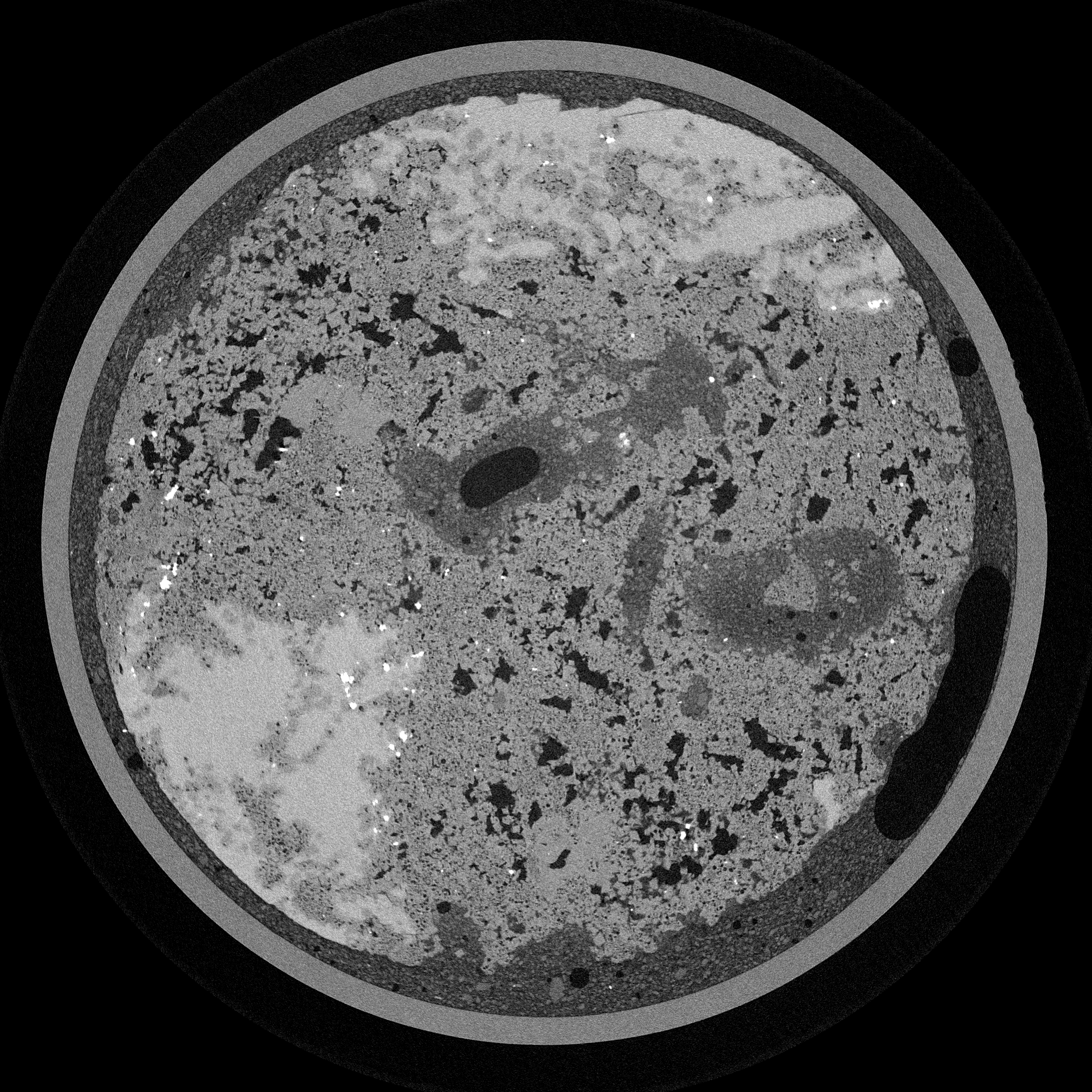}{7}{0.220,0.560}{0.290,0.630}{0.490,0.025}
    \vspace*{-0.5cm}%
    \label{fig:rock_slices_DH_Z}%
    }%
    \subfloat[SFT; GBC; z-slice]{
    \hspace*{-0.2cm}
    \zoominset{0.322\linewidth}{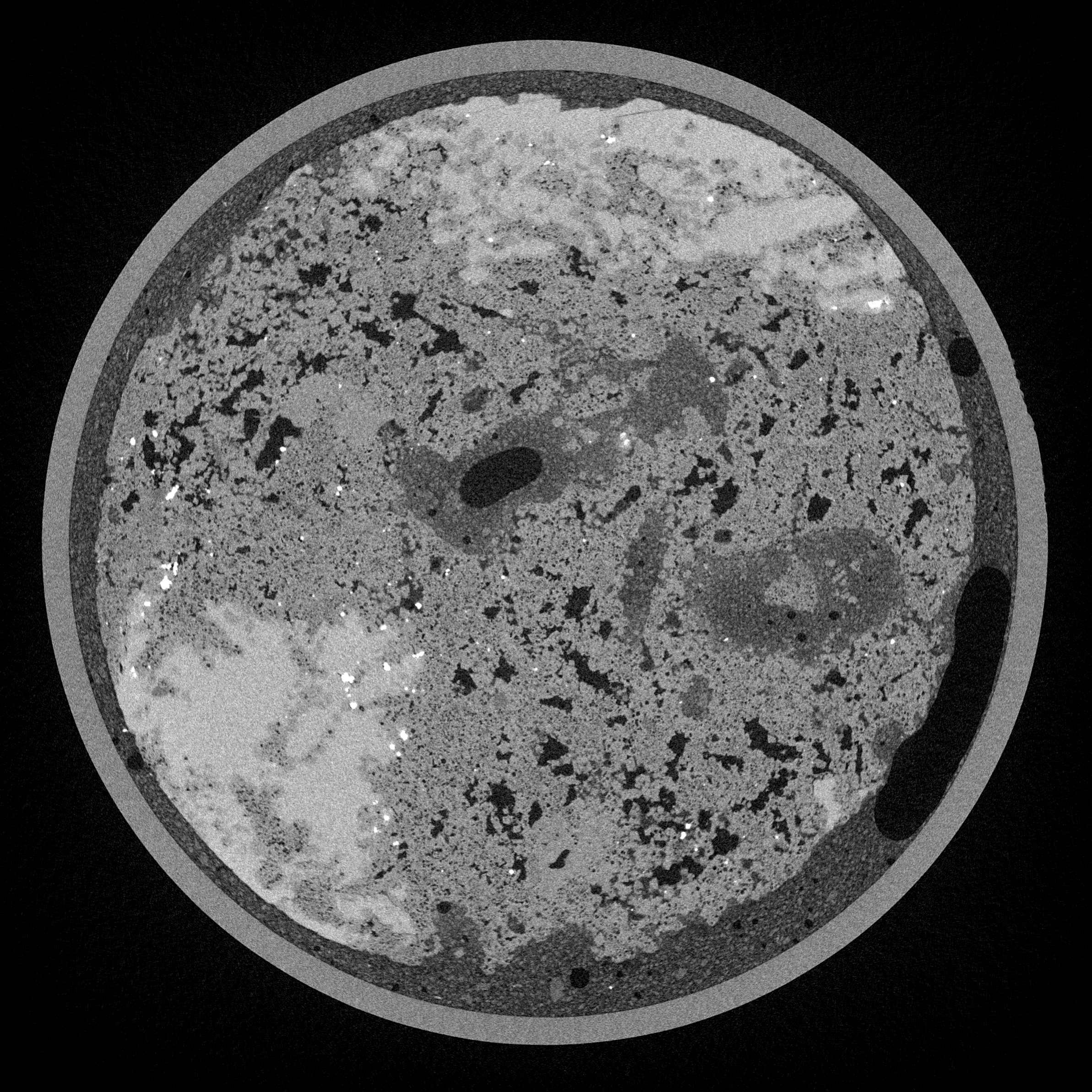}{7}{0.220,0.560}{0.290,0.630}{0.490,0.025}
    \vspace*{-0.5cm}
    }%
    \subfloat[SFT; iterative; z-slice]{
    \hspace*{-0.2cm}
    \zoominset{0.322\linewidth}{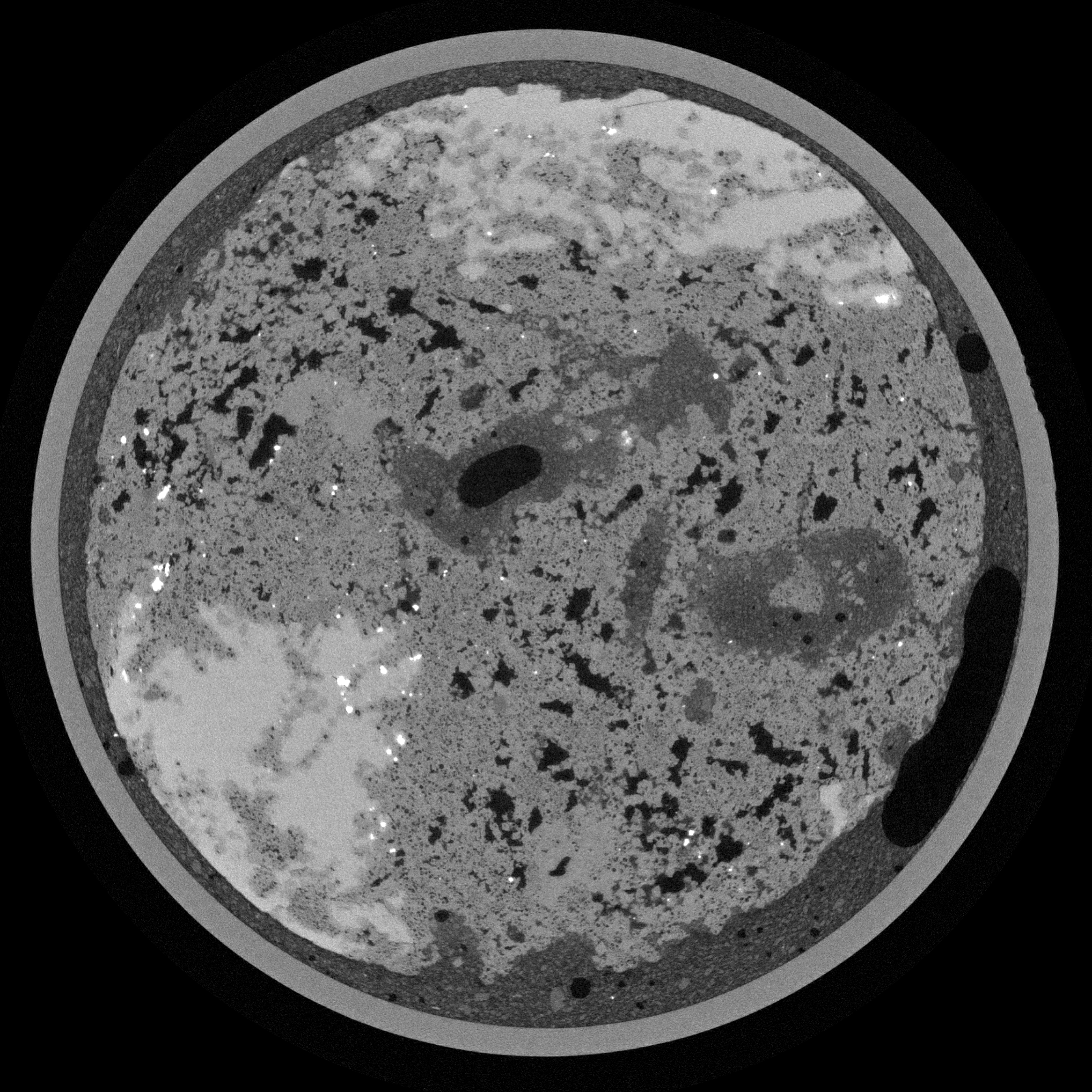}{7}{0.214,0.561}{0.284,0.631}{0.490,0.025}%
    \vspace*{-0.5cm}
    }%
    \end{minipage}%
    \newline%
\caption{Slices of the tomograms reconstructed from experimental data. \textbf{Left:} a reconstruction from double-helical data (9244 source points), using a Katsevich-like filtered-backprojection algorithm. \textbf{Middle:} a reconstruction from the SFT trajectory data (5754 source points), using Global Backprojection-Convolution. \textbf{Right:} a reconstruction from the same SFT trajectory data, using the iterative multi-scale method described in \cite{myers2016rapidly}. \textbf{All:} plotting scale: attenuation coefficient of $0$ (black) to $0.25\;\mathrm{mm}^{-1}$ (white). Note that the GBC images include a significant DC correction of $+0.0355 \; \mathrm{mm}^{-1}$.}
    \label{fig:rock_slices}
\end{centering}
\end{figure*}

See Figure~\ref{fig:rock_slices} for slices of the reconstructions. The GBC reconstruction includes significant low-frequency errors due to the object extending axially beyond the source cylinder. The DC component of this error is straightforward to correct manually, and we have done so for the figure, to aid comparison.
An exact match is not possible between the slices from the double-helix reconstruction and those from the SFT reconstructions, because these are from separate scans and the precision of instrumentation is limited. We  have manually selected the closest matching reconstruction slices.

\subsubsection{Discussion}
Compared with the double-helix (DH) reconstruction, the GBC and iterative reconstructions achieved greater resolution isotropy. This is particularly visible in the zoomed inset of the Figure~\ref{fig:rock_slices_DH_Z}, where the metal inclusion (bright spot) has smeared and produced an anisotropic artefact. This difference was expected, and is understood to be due to the use of the DH vs. SFT scanning trajectories \cite{varslot2012considerations}.

The GBC reconstruction is nearly as sharp as the DH reconstruction and has a similar noise profile, despite using almost half the number of source points. The iterative reconstruction has visibly less noise; this is likely due to a combination of a lack of iterations (i.e. unintentional regularisation by early-stopping of an iterative method) and the filtering and regularisations applied as part of the algorithm in \cite{myers2016rapidly}.
The iterative reconstruction also features `bloom' artefacts around the metal inclusions, whereas the GBC reconstruction has better resolved the boundaries of these metal inclusions.

Although GBC is not correctly applicable to long objects, it appears in this instance to have sufficiently handled the low-frequency errors introduced by the long object problem, via the low-pad correction described in \S\ref{sec:lowpad_correction}.

If the iterative method were allowed to iterate indefinitely, then it would likely converge on a superior reconstruction to GBC (i.e. similarly sharp, but less noisy), because it is based on a maximum-likelihood method that accounts for the photon statistics. The major benefit of GBC is that it is a direct reconstruction method requiring only a single weighted backprojection and a volume convolution. The iterative method requires at least one projection and one backprojection per iteration at full scale, and there were two such iterations performed for this comparison. GBC could be used to initialise the multi-scale iterative method, thereby obviating the need for the `multi-scale' complexity of method, allowing it to begin iterating at full resolution immediately.

\section{Conclusions}\label{sec:conclusion}
We have introduced what we believe to be the first practical algorithm in transmission cone-beam computed tomography (CBCT) to directly reconstruct a tomogram from data acquired with a multidimensional source locus. We refer to this algorithm, and the method more broadly, as `global backprojection-convolution' (GBC). The theory generalises previously established `deconvolution algorithms' in the PET literature, e.g. \cite{colsher1980fully,schorr1983general,defrise1995}, and extends the idea for use in CBCT in the native data format, without the need for data rebinning.

The cylindrical locus served as our prototype and proof-of-concept. 
The regularised cylinder-based algorithm given in this article may, in principle, be used on any cylinder-filling transmission source locus---the distribution of points need only be uniform (i.e. equal number of source points per unit area). Indeed, we demonstrated reconstruction from two such loci. (The low-discrepancy sequence and the `space-filling trajectory' from \cite{kingston2018space}.)
Cylindrical loci have advantages over helical or helix-like loci---these were described in the introduction to this article---and are already readily accessible to most existing helical CT scanners operating in a `step-and-shoot' mode. The improved resolution isotropy is visible in the experimental reconstruction slices.
The cylindrical algorithm we developed is fast and precise, as evidenced by the reconstructions on simulated data. Only a single backprojection is required, followed by a volume convolution. No forward projection is required. The reconstruction on experimental data suggests that the algorithm is robust to basic experimental sources of error, and to mild violation of validity assumptions. The algorithm can be applied to long objects, if one is willing to accept low-frequency errors.
The discretisation regularisations developed in this article seem sufficient to enable GBC to be practically applied to non-ideal datasets.

The theory of GBC and the discretisation regularisations presented in this article can be generalised to other multidimensional loci, such as the sphere. Such generalisations may serve as a basis for future research. 

It would be desirable to more quantitatively characterise the error of the discretisation regularisations. An improved understanding of the relationship between reconstruction error and the choice of hyperparameters (e.g. padding factors) would aid in the appropriate choice of these hyperparameters. Similarly, a more careful analysis of point sampling patterns on multidimensional loci could improve stability and noise performance. The LDS trajectory used in the simulation section of this article was convenient for programmatic/computational reasons, but is not advised for actual scanning due to the impractical scanning time associated with following the LDS sequence directly. 
A more quantitative characterisation of the low-frequency error would allow us to understand what errors are included when this method is applied to long objects.

The reconstructions in this article may be reproduced with e.g. \texttt{aprict: v0.1.4} \cite{aprict-v0.1.4}. The experimental data and reconstruction scripts are available at \cite{exp_data}.

\section{Future Work}
 We have also applied this reconstruction procedure successfully to the spherical geometry depicted in Fig.~\ref{fig:trajectories_lds_sphere}. This and other geometries may be the subject of future work, as may numerical techniques for rapidly computing the backprojection weights and deconvolution filter for less symmetric geometries, wherein closed-form expressions are not available. It is also possible to alter the weighting functions so that opposing rays are weighted differently, in a position-dependent manner.

\section*{Acknowledgments}\label{sec:acknowledgments}
The author acknowledges funding through an Australian Government Research Training Program scholarship. 
The author extends their gratitude to Robert Middleton and the Australian National University CT Lab, who provided the experimental scanning data and the double-helix and iterative reconstructions.
Thanks to Fr\'ed\'eric Noo for referral to related literature in the field of PET imaging, and for discussion of Schwartz spaces. 
Thanks also to Ayush Bhandari for helpful remarks on the theoretical grounding of the work.

{\appendices
\section{}\label{app:appendix}

\subsection{Exact softening scheme}
\label{app:exact_soften}
In the body of the article, we mentioned that it is necessary to anti-alias or `soften' the vertical extremities of the backprojected radiographs. The purpose of this softening is to ameliorate discretisation artefacts that otherwise surface due to aliasing. However, such softening alters the convolution, requiring that frequency component $\xi$ be multiplied by the coefficient $|\xi|/\mathcal I[f](\hat{\xi})$ where the Funk transform $\mathcal I[f](\hat{\xi})$ differs from its nominal value in \eqref{eq:cyl_eqs_I} due to the effect of the softening on the backprojection weighting term $W(x, r)$.

In this appendix, we describe a softening formula that admits a closed-form solution for the Funk transform. This means that the resultant combination of softening and convolution are theoretically exact, yet they do not require numerically evaluating integrals.

For brevity, we do not include a derivation, but we will describe the approach. We arrived at this formula by first writing an explicit formula for the Funk transform associated with $\xi$, and then noticing that the integral could be evaluated exactly if the softening function $s(\vartheta_v, \theta_{\text{soft}} ; \theta_{\text{el}})$ were a polynomial in $\sin |\theta_{\text{el}}|$. We first chose the unique affine transformation of $\sin|\theta_{\text{el}}|$, termed $x$, that would be equal to $0$ at $|\theta_{\text{el}}| = \vartheta_v$ and be equal to $1$ at $|\theta_{\text{el}}| = \vartheta_v - \theta_{\text{soft}}$. Then we swapped out the expression $x$ for $3x^2 - 2x^3$ in order to make the softening function differentiable.

Here is the exact combination of softening + convolution that we devised. 
The softening function is given by:
\begin{align}\label{eq:exact_softening_function}
    s(\vartheta_v, \theta_{\text{soft}} ; \theta_{\text{el}}) 
    = \begin{cases}
    0 & \text{if} \qquad |\theta_{\text{el}}| \geq \vartheta_v \\
    1 & \text{if} \qquad |\theta_{\text{el}}| \leq \vartheta_v - \theta_{\text{soft}} \\
    3x^2 - 2x^3 & \text{otherwise} 
    \end{cases} \\
    \shortintertext{\text{where} $x = \gamma \sin |\theta_{\text{el}}| - \chi \, ,$ }
    \text{and } \gamma = \left( \sin\left(\vartheta_v - \theta_{\text{soft}} \right) - \sin(\vartheta_v)\right)^{-1} , \quad
    \chi = \gamma \sin(\vartheta_v) . \nonumber
\end{align}
The associated convolution amounts to a multiplication of each Fourier component $\xi$, with polar angle $\theta_{\xi}$ measured from the axis of the cylinder, by its magnitude $|\xi|$ and by the reciprocal of the following exact formula for the Funk transform that replaces $\mathcal I[f](\hat{\xi})$:
\begin{subequations}\label{eq:closed_form_integral}
\begin{align}
    &G(\vartheta_v, \theta_{\text{soft}} ; \xi) =
    a + b \sin\left(\theta_{\xi}\right) + c \sin\left(\theta_{\xi}\right)^2 + d \sin\left(\theta_{\xi}\right)^3 
    \intertext{\text{where}}
    a &= 4 A \chi^2 (3 + 2 \chi) + 4 \arcsin(\alpha)\\
    b &= 24 \gamma \chi (1 + \chi) (B - C) \\
    c &= 6 \gamma^2 (1 + 2 \chi) (A - \beta B + \alpha C)\\
    d &= -(2/3) \gamma^3 \left( \big. 
        D - 9(B - C)
    \right)
    \intertext{\text{where in turn}}
    \alpha &= \sin(\vartheta_v - \theta_{\text{soft}}) / \max\left\lbrace \sin(\vartheta_v - \theta_{\text{soft}}), \; \sin \theta_{\xi} \right\rbrace \\
    \beta &= \sin(\vartheta_v) / \max\left\lbrace \sin(\vartheta_v), \; \sin \theta_{\xi} \right\rbrace \\
    A &= \arcsin(\beta) - \arcsin(\alpha) \\
    B &= \sqrt{1 - \beta^2} \\
    C &= \sqrt{1 - \alpha^2} \\
    D &= \cos(3\arcsin(\beta)) - \cos(3\arcsin(\alpha)) \, .
\end{align}
\end{subequations}
Note that $\arccos, \arcsin : [0,1] \rightarrow [0, \tfrac \pi 2]$, and $\theta_{\xi}, 2\vartheta_v \in [0, \pi]$.

\subsection{Discrete Convolution}
\label{app:discrete_fourier_filtration}
In \S\ref{sec:discretising_fourier_filtration} we mentioned that the convolution must be modified before it can be correctly applied to the conventional Discrete Fourier Transformation (DFT). In this appendix, we explain our thinking on this subject, and try to justify the approximation given in that section of the article.

The convolution formula \eqref{eq:deconv_filter_int} was derived in \S\ref{sec:inversion_theory} 
from a consideration of the tomogram volume as a function on continuous space $\mathbb R^3$. In practice, we represent the volume discretely, as a 3D grid of voxels. It is nonsensical to apply the continuous convolution formula (or, for example, the ramp filter) to a discrete volume. The natural resolution to this issue is to consider the discrete volume as a kind of `compressed encoding' of a continuous volume. 

Denote the locations of the voxel centres within $\mathbb R^3$ by $\mathbf p_i$. We write the discrete volume attenuation coefficients as $f(\mathbf p_i) = f_i$. We imagine that the discrete volume encodes the continuous volume by a convolution of a 3-dimensional `comb' of Dirac-delta distributions:
\begin{equation}
    f = g * \sum_i f_i \delta_{\mathbf p_i} \, ,
\end{equation}
where $\delta_{\mathbf p_i}$ is a 3-dimensional Dirac-delta distribution with its peak at $\mathbf p_i$, and $g$ is the convolution kernel, and $*$ is the convolution operator. For example, the kernels associated with the zeroth-order and first-order (trilinear) encodings mentioned above are given respectively by
\begin{align}
    g_0(\mathbf v) &= \begin{cases} 
        \max\{|v_x|, |v_y|, |v_z|\} > 1/2 :& 0 \\
        \text{else} :& 1
    \end{cases} \\
    g_1(\mathbf v) &= \prod_{i \in \{x,y,z\}} \max\left\lbrace 0, \; 1 - |v_i| \right\rbrace \, .
\end{align}

The Fourier transform of the continuous volume is, by the convolution theorem, given by the pointwise-product of the Fourier transforms of the two functions that are convolved:
\begin{align}
    \tilde{f} &= \mathcal F \left[ f \right] 
        = \tilde g \times \left( \sum_i f(\mathbf p_i) \tilde \delta_{\mathbf p_i} \right) \\
    \tilde{f}(\mathbf k) &= \tilde g(\mathbf k) \left(\sum_i f_i e^{-i \mathbf k \cdot \mathbf p_i} \right)  \, ,
\end{align}
where the pointwise-product of functions is defined by $(a \times b)(x) = a(x) b(x)$. To perform the convolution in the continuous domain, we multiply $\tilde f$ by the appropriate transfer function (e.g. $|\mathbf k / (2 \pi)|/\mathcal I[f](\mathbf k / |\mathbf k|)$, or $|\mathbf k / (2 \pi)|$ for the ramp filter), which we denote $\tilde t(\mathbf k)$. The Fourier transform of the filtered continuous volume is 
\begin{equation*}
    \tilde t(\mathbf k) \tilde g(\mathbf k) \left(\sum_i f_i e^{-i \mathbf k \cdot \mathbf p_i} \right) 
\end{equation*}
Finally, to recover a discrete encoding of the resultant volume, we resample the resulting function at points $\mathbf p_i$, after optionally convolving it once more with a kernel $h$. (For example, with $h = g_0$, the voxels would inherit the integrated attenuation within their voxel. With $h=1$, i.e. $h$ absent, the voxels would inherit the value sampled from their centre.) Whatever the choice of $g, h$ and whatever the transfer function $\tilde t$, the resulting expression for the filtered function in the continuous domain is
\begin{equation*}
    \tilde y(\mathbf k) = \tilde t(\mathbf k) \tilde g(\mathbf k) \tilde h(\mathbf k) \left(\sum_i f_i e^{-i \mathbf k \cdot \mathbf p_i} \right) \, ,
\end{equation*}
By design, the discrete samplings $y(\mathbf p_i) = y_i$ will be our filtered discrete volume.
Computing the inverse Fourier transformation at point $\mathbf p_j$ on the discretised volume, we find
\begin{align*}
    y_j &= \iiint \mathrm d^3 \mathbf k \, \tilde t(\mathbf k) \tilde g(\mathbf k) \tilde h(\mathbf k) \left(\sum_i f_i e^{i \mathbf k \cdot (\mathbf p_j - \mathbf p_i)} \right) \\
        &= \sum_i f_i \iiint \mathrm d^3 \mathbf k \, \tilde t(\mathbf k) \tilde g(\mathbf k) \tilde h(\mathbf k) e^{i \mathbf k \cdot (\mathbf p_j - \mathbf p_i)} \\
        &= \sum_i f_i \mathcal F^{-1} \left[ \tilde t(\mathbf k) \tilde g(\mathbf k) \tilde h(\mathbf k) \right](\mathbf p_j - \mathbf p_i) \, .
\end{align*}
This is a discrete convolution between $f$ and the discrete samplings of the continuous kernel $K = \mathcal F^{-1} \left[ \tilde t \times \tilde g \times \tilde h(\mathbf k) \right] $.

The extent of $K$ may be infinite, such that the discrete convolution cannot be computed in a finite number of operations. However, in practice the kernel $K$ will drop off quickly from the origin, as is our case. It is convenient for us to implement the discrete convolution as a multiplication between components of the conventional Discrete Fourier Transformations (using the discrete convolution theorem; note that this produces a cyclic convolution). 
The formula for the convolution $y$ of the discrete volume $f$ is
\begin{align}\label{eq:exact_discrete_convolution}
    y &= \mathrm{DFT}^{-1} \left[ 
        \mathrm{DFT}\left[
            C
        \right] \times \mathrm{DFT}[f]
    \right] \\
    \text{where} \quad C &= \text{discretely sampled} \; \mathcal F^{-1} \left[ \tilde t \times \tilde g \times \tilde h \right] \, ,\nonumber
\end{align}
and where the symbol $\times$ represents a pointwise multiplication between two discrete Fourier transforms, and the operator $\mathrm{DFT}$ is the conventional discrete Fourier transform,

Computing the discretely sampled $C$ is difficult due to the 3-dimensional integration over $\mathbf k$, with no immediately obvious symmetries that can be exploited to perform integration analytically, even partially. As an approximation, we may sample the $\tilde t(\mathbf k) \tilde g(\mathbf k) \tilde h(\mathbf k)$ discretely, yielding the approximation:
\begin{align}\label{eq:approx_discrete_convolution}
    y &= \mathrm{DFT}^{-1} \left[ 
        \tilde t \times \tilde g \times \tilde h \times \mathrm{DFT}[f]
    \right] \, .
\end{align}
Ideally, the integrals would be computed analytically to produce an exact discrete convolution from the continuous transfer function $\tilde t$ (c.f. \cite{crawford1991}). When those integrals don't have closed form, the above approximation may be superior to the na\"ive approach of sampling $\tilde t$ discretely.

We have determined emprically that choosing $g = g_0$ and $h=1$ yields significantly improved reconstructions compared with sampling $\tilde t$ discretely: high-frequency artefacts are greatly suppressed, with no obvious introduction of new artefacts. This amounts to the following modification to the na\"ive discrete formula for the convolution:
\begin{align*}
    &\text{From:}& y &= \mathrm{DFT}^{-1} \left[ \tilde t \times \mathrm{DFT}[f] \right] \\
    &\text{to:}& \quad y &= \mathrm{DFT}^{-1} \left[ \tilde g_0 \times \tilde t \times \mathrm{DFT}[f] \right] \\
    &\text{where:}& \quad \tilde g_0(\mathbf k) &= 
        \mathrm{sinc}\!\left(l k_x/2\right)
        \mathrm{sinc}\!\left(l k_y/2\right)
        \mathrm{sinc}\!\left(l k_z/2\right) \, ,
\end{align*}
where $l$ is the side length of the cubic voxels, and $\mathrm{sinc}$ is 
\begin{equation}\label{eq:sinc_function}
    \mathrm{sinc}(x) = \begin{cases}
        x = 0: & 1 \\
        x \neq 0: & \frac{\sin(x)}{x}
    \end{cases} \, .
\end{equation}

\subsection{Source point number for data sufficiency with the cylindrical source locus}
\label{sec:appendix_sufficiency}

We have a \emph{rough} estimate of the sufficient number of source points that is based on the Crowther criterion \cite{crowther1970reconstruction}, assuming that the source point distribution is isotropic. 

For smaller vertical cone angles, the Fourier frequencies that are sampled most sparsely are those in the \emph{lateral plane}. As an approximation, we consider the number of backprojections received by a volume point from source points at all heights, at some azimuthal angle $\phi$. The worst-case scenario (least source points per radian in $\phi$) is at any point on the edge of the object support radius $r$ looking in the direction tangent to the circle of radius $r$. 
We require the projection density to be such that this worst-case scenario still yields the minimum number of source points per radian. According to the Crowther criterion, that number is $D/4$ if we assume that the source points are evenly spaced, where $D$ is the width---in voxels---of the object support diameter $2r$. However, we observe that many source points are oppositely oriented (or close to it) by coincidence, and so correspond to the same \emph{view}. For that reason, the required number must be doubled to ensure sufficiency, i.e. we require $D/2$ source points per radian.
According to this reasoning, the required density $\mu$ of source points per unit area of the source cylinder with radius $R$ is:
\begin{equation}
    \mu \geq \frac{1}{4Rw} \frac r R \Big/ \left( \Big. \tfrac 1 2 \tan(\vartheta_v) \sqrt{1-(r/R)^2} \right) \, , 
\end{equation}
where $w$ is the width of a voxel, $r$ is the object support radius, $R$ is the source cylinder radius, and $\vartheta_v$ is as described in \S\ref{sec:specialisation_to_cylinder}. 

We find that for larger cone angles, it is in fact the frequencies close to the poles $\theta_{\mathbf k} = 0$ and $\theta_{\mathbf k} = \pi$ that are least sampled, and the worst-case scenario is at points in the volume that coincide with the cylinder axis. In a similar computation to the lateral one, we find the number of source points swept out per radian of tilt in a horizontal plane, as it tilts around the axial point. We find the bound
\begin{equation}
    \mu \geq \frac{1}{4Rw} \frac{r}{R} \, .
\end{equation}
Combining the bounds, our estimate for the required number of source points per voxel-height of the source cylinder, $\Lambda_z = \mu 2 \pi R w$,  is
\begin{equation}\label{eq:data_sufficiency_01}
    \Lambda_z \geq \frac \pi 2 \frac{r}{R} \max\left\lbrace
        1 \, , \quad
        \left( \Big. \tfrac 1 2 \tan(\vartheta_v) \sqrt{1-(r/R)^2} \right)^{-1}   
    \right\rbrace \, .
\end{equation}

Next, we simplify to the case where the reconstruction support radius $r$ is taken as its maximum value $R \sin(\vartheta_h)$, i.e., we assume that the object fills the horizontal extent of the detector. Then the required number of source points per voxel-height of the source cylinder is
\begin{equation}\label{eq:data_sufficiency_approx}
    \Lambda_z \geq \pi \max \left\lbrace 
        \tfrac 1 2 \sin(\vartheta_h) \, , \quad
         \frac{\tan(\vartheta_h)}{\tan(\vartheta_v)}
    \right\rbrace \, ,
\end{equation}
\begin{equation}
    \text{i.e. }
    \Lambda_z \geq \pi \max \left\lbrace 
        \frac{W}{4 L \sqrt{1 + \frac{W^2}{4L^2}}} 
        \, , \quad
        \frac W H \sqrt{1 + \frac{W^2}{4L^2}}
    \right\rbrace ,
\end{equation}
with $\vartheta_h$ as described in \S\ref{sec:specialisation_to_cylinder}, $W$ the detector width, $H$ the detector height, and $L$ the distance from the source point to the detector. 
If we make the additional simplifying assumption that the detector is square, $W = H$, then the lateral bound is sharper, and \eqref{eq:data_sufficiency_approx} reduces to
\begin{equation}\label{eq:data_sufficiency_approx_02}
    \Lambda_z \geq \pi \sec(\vartheta_h) = \pi \sqrt{1 + \frac{W^2}{4L^2}}\, .
\end{equation}
}

\bibliographystyle{IEEEtran}
\bibliography{refs.bib}

\vfill

\end{document}